\documentclass[usenatbib]{mnras}
\usepackage[english]{babel}

\usepackage[utf8x]{inputenc}
\usepackage{amsfonts}
\usepackage{amsmath}
\usepackage{amssymb}
\usepackage{xcolor}
\usepackage{enumitem}
\usepackage{graphicx}
\usepackage{gensymb}
\usepackage{fancyhdr}
\usepackage{times}
\usepackage{widetext}
\usepackage{ulem}
\usepackage[caption = false]{subfig}
\usepackage{lineno}
\newcommand{\commentOut}[1]{}

\newcommand{\D}{\mathrm{d}}

\newcommand{\crit}{\ensuremath{_\mathrm{cr}}}

\graphicspath{{./Figures/}} 

\begin{document}

\title[Baryon scaling relations of galaxy groups in GAMA and BAHAMAS]{Multi-wavelength scaling relations in galaxy groups: a detailed comparison of GAMA and KiDS observations to BAHAMAS simulations}

\author[Jakobs et al.]%
{Arthur Jakobs$^{1,2}$,
Massimo Viola$^2$, Ian McCarthy$^3$, Ludovic van Waerbeke$^1$\thanks{waerbeke@phas.ubc.ca}, 
\newauthor{Henk Hoekstra$^2$, Aaron Robotham$^4$, Gary Hinshaw$^1$, Alireza Hojjati$^1$, Hideki Tanimura$^1$,}
\newauthor{Tilman Tröster$^1$, Ivan Baldry$^3$, Catherine Heymans$^5$, Hendrik Hildebrandt$^6$,} 
\newauthor{Konrad Kuijken$^2$, Peder Norberg$^7$, Joop Schaye$^2$, Cristóbal Sif\'on$^8$, Edo van Uitert$^9$,} \newauthor{Edwin Valentijn$^{10}$, Gijs Verdoes Kleijn$^{10}$, Lingyu Wang$^{10,11}$}\\
$^1$Department of Physics and Astronomy, University of British Columbia, Vancouver, V6T 1Z1, BC, Canada\\
$^2$Leiden Observatory, Leiden University, PO Box 9513, NL-2300 RA Leiden, Netherlands\\
$^3$Astrophysics Research Institute, Liverpool John Moores University, 146 Brownlow Hill, Liverpool L3 5RF\\
$^4$International Centre for Radio Astronomy Research (ICRAR), The University of Western Australia, 35 Stirling Highway, Crawley, WA 6009, Australia\\
$^5$Scottish Universities Physics Alliance, Institute for Astronomy,University of Edinburgh, Royal Observatory, Blackford Hill,Edinburgh EH9 3HJ, UK\\
$^6$ Argelander-Institut für Astronomie, Auf dem Hügel 71, 53121 Bonn, Germany\\
$^7$ ICC \& CEA, Department of Physics, Durham University, South Road, Durham DH1 3LE, UK \\
$^8$ Department of Astrophysical Sciences, Peyton Hall, Princeton University, Princeton, NJ 08544, USA\\
$^9$Department of Physics and Astronomy, University College London, Gower Street, London WC1E 6BT, UK\\
$^{10}$Kapteyn Astronomical Institute, University of Groningen, PO Box 800, 9700 AV Groningen, Netherlands\\
$^{11}$SRON Netherlands Institute for Space Research, Landleven 12, NL-9747 AD Groningen, Netherlands}

\maketitle
\begin{abstract}
We study the scaling relations between the baryonic content and total mass of groups of galaxies, as these systems provide a unique way to examine the role of non-gravitational processes in structure formation.  Using Planck and ROSAT data, we conduct detailed comparisons of the stacked thermal Sunyaev-Zel'dovich (tSZ) and X-ray scaling relations of galaxy groups found in the Galaxy And Mass Assembly (GAMA) survey and the BAHAMAS hydrodynamical simulation.  We use weak gravitational lensing data from the Kilo Degree Survey (KiDS) to determine the average halo mass of the studied systems. We analyse the simulation in the same way, using realistic  weak lensing, X-ray, and tSZ synthetic observations.  Furthermore, to keep selection biases under control, we employ exactly the same galaxy selection and group identification procedures to the observations and simulation.  Applying this comparison, we find that the simulations reproduce the richness, size, and stellar mass functions of GAMA groups, as well as the stacked weak lensing and tSZ signals in bins of group stellar mass.  However, the simulations predict X-ray luminosities that are higher than observed for this optically-selected group sample.  As the same simulations were previously shown to match the luminosities of X-ray-selected groups, this suggests that X-ray-selected systems may form a biased subset. Finally, we demonstrate that our observational processing of the X-ray and tSZ signals is free of significant biases. We find that our optical group selection procedure has, however, some room for improvement.
\end{abstract} 


\section{Introduction}\label{sec:introduction}
In the currently favoured $\Lambda$CDM model, structure forms hierarchically from small density fluctuations that are observed as minute temperature fluctuations in the cosmic microwave background \citep[CMB;][]{Planck16}. Although dark matter is the driving force behind the formation of the large-scale structure, it is nonetheless crucial to understand the distribution and observable properties of the baryonic matter: while it may not play a major role in structure formation, it does provide the link between the  observable universe and the underlying distribution of matter.  Furthermore, to do so-called `precision cosmology' with large-scale structure,  an accurate characterisation of baryonic effects on the total mass distribution is required (e.g., \citealt{Semboloni11,VanDaalen2011}).

Individual galaxies may be viewed as the main building blocks of large-scale structure, but the continuous accretion of smaller structures into larger ones results in galaxy groups being the most common environment in which galaxies are found.  Bridging the gap between field galaxies and massive clusters, galaxy groups fill in an important phase of structure formation and it is thought that most galaxies are either part of a group or have been part of a group in the past \citep{Eke2004}.  Groups have not been studied as extensively as clusters of galaxies or galaxies themselves. This is likely because galaxy groups are difficult to identify observationally, given the relatively low number of galaxies they comprise and their low contrast against the background.  Only recently, with the advent of large spectroscopic surveys, have substantial samples of groups become available.  

The halo mass, a key quantity, can only be measured indirectly for individual groups.  Although they can be studied using deep X-ray observations \citep[e.g.][]{Sun2008}, a simple interpretation of such results may be affected by non-gravitational physical processes, such as active galactic nuclei (AGN) and feedback linked to star formation and supernovae. These processes have a strong effect on the distribution of matter in groups and in particular baryons \citep{Fabjan2010, McCarthy2010,LeBrun2014,Velliscig2014}, because the gravitational binding energy of groups is not as large as that of galaxy clusters, where they don't play a major role in their mass content.  

The gravitational potential wells of galaxy groups are deep enough to retain some fraction of the baryons, so the main effect of the various feedback processes is to change the distribution of the different components and therewith the correlations between the various observable properties. These scaling relations are the result of the various processes that govern the formation of galaxy groups. This makes them ideal targets for studying the effect feedback processes have on the matter distribution.  Hydrodynamical simulations have shown how various feedback processes can affect the distribution of baryonic and non-baryonic matter at all mass scales (e.g. \citealt{Mummery2017}). This effect has recently also been measured in the cross-correlation between baryonic and non-baryonic probes such as the thermal Sunyaev-Zel'dovich (tSZ) signal and gravitational lensing \citep{VanWaerbeke2014,Hill2014,Battaglia2015,Hojjati2015,Hojjati2016}. A better understanding of galaxy group scaling relations can  help to promote groups as a robust statistical cosmological probe and shed light on the underlying mass scale. While a better understanding of feedback processes also helps to improve constraints from cosmic shear studies \citep{Semboloni11,Semboloni13}.

Detailed multi-wavelength studies of individual groups provide key information on the scatter in scaling relations, but are expensive. Fortunately, a great deal can be learned by characterising their average scaling relations, which can be obtained by considering the properties of ensembles of groups  (i.e. stacking signals of subsets selected by some observable, such as stellar mass, etc.). Thanks to wide-area surveys in X-ray, optical and millimetre wavelengths, such scaling relations can now be measured with good precision.  However, in stacking analyses object selection becomes particularly important, as the interpretation relies on an understanding of the underlying population.  For instance, X-ray-selected samples may be biased if they preferentially pick out X-ray luminous/gas-rich systems. The best strategy is then to select a clean sample using a different (independent) indicator and stack the observables of interest for the entire sample. For instance, \citet{Anderson2015a} argued that samples based on optical properties are not prone to the X-ray selection bias. 

\citet{Anderson2015a} used a sample of 'locally brightest galaxies' (LBG) defined by \citet{Planck2013} and measured the stacked X-ray luminosity,
whereas \citet{Planck2013} studied the stacked tSZ signal.  The rationale for using LBGs is that they typically correspond to the central galaxy in a dark matter halo. These studies bin their sample in LBG stellar mass and use this as a proxy for halo mass, using the stellar-to-halo mass relation predicted by the semi-analytic model of \citet{Guo2011}. \cite{Wang2016} have done a recalibration of the \citet{Anderson2015a} result using gravitational lensing, in order to eliminate the model dependence linking the central galaxy luminosity to its parent halo mass.   The resulting X-ray luminosity-mass and tSZ-mass scaling relations may, however, be difficult to interpret if there is significant scatter in the correlations between the different observables used.  This is where realistic numerical hydrodynamical simulations can be helpful, as they offer a way to interpret the observations and study the effect of these physical processes on the matter distribution in the Universe.

In this paper we study the X-ray and tSZ effect scaling relations using a large sample of galaxy groups from the Galaxy And Mass Assembly (GAMA) survey \citep{Driver2011}, a large spectroscopic survey that is ideally suited to identify galaxy groups.  In contrast to \citet{Anderson2015a}, we determine the halo masses using stacked weak gravitational lensing measurements from the Kilo-Degree Survey \citep[KiDS;]{Kuijken2015, Hildebrandt2016a, deJong2017}. We use the X-ray measurements from the ROSAT All Sky Survey X-ray data \citep{Voges1992} and the Planck Compton-$y$-map \citep{Planck2015} for the tSZ measurements. The groups are identified using (a modified version of) the Friends-of-Friends group finding algorithm employed for the GAMA survey \citep{Robotham2011}. Crucially, we apply the same algorithm to the BAryons and HAloes of MAssive Systems simulations \citep[BAHAMAS;][]{McCarthy2017}, so that we obtain two identically-selected group samples.  We use the integrated stellar mass of the groups (which is a proxy for the total halo mass) to stack the other observables.

This paper is organised as follows: In Section \ref{sec:data} we introduce the GAMA data and describe the BAHAMAS simulations. We discuss the group selection process and final samples in Section \ref{sec:FoF}. The stacking procedures and relevant datasets  are introduced in Section \ref{sec:DataStacking}. We present our main results in Section \ref{Results} and discuss the impact of selection effects on these results in Section \ref{fragmentation}.  Finally, we discuss and summarise our results in Section \ref{Discussion}. We note that throughout this paper we use $\log = \log_{10}$. 

\section{Observed and simulated data}

\label{sec:data}
\begin{figure*}
\includegraphics[width=\textwidth,trim=+5cm 0 1cm -1cm]{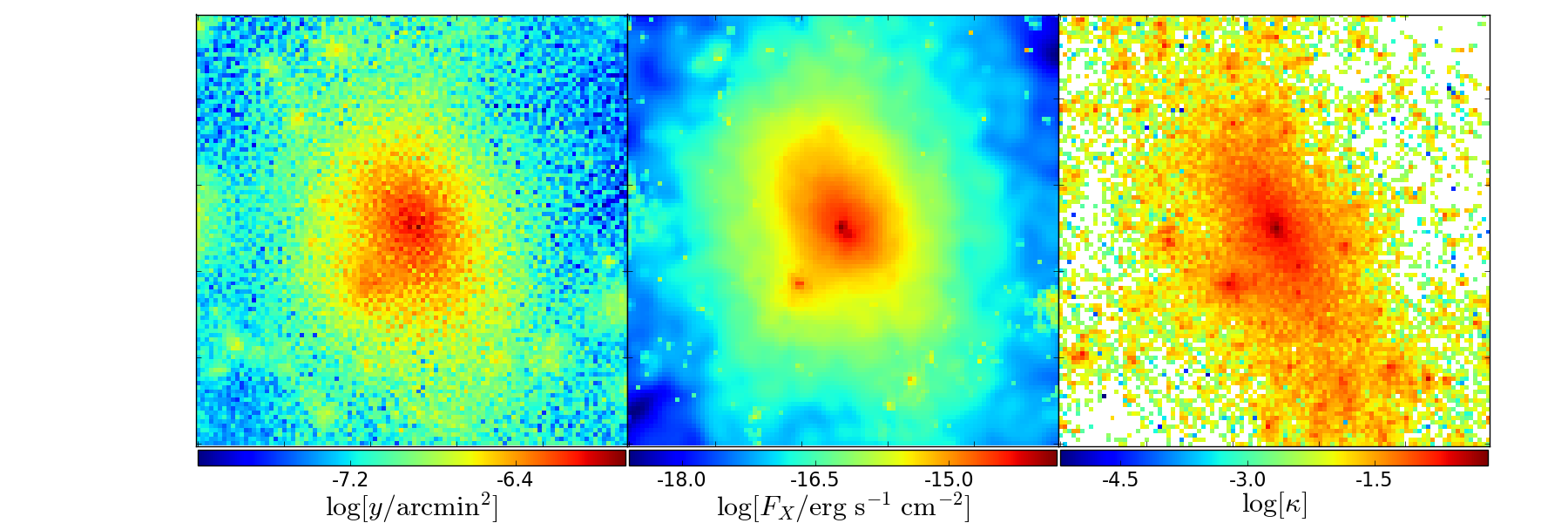}
\caption{The three panels show $17'\times17'$ cut-outs from the BAHAMAS light cone maps. From left to right we show
the tSZ signal ($y$), the X-ray flux ($F_{\rm X}$) and the lensing convergence ($\kappa$), centred on the most massive cluster, $\log[M_{500}/(h^{-1}M_\odot)] = 14.27$. The X-ray map is in the 0.5-2.4 keV energy band and the $\kappa$ map was generated using the KiDS source redshift distribution.}\label{Simsnapshots}
\end{figure*}

\subsection{The GAMA group sample}\label{GAMA}

In contrast to clusters of galaxies, galaxy groups are more difficult to identify using optical imaging data. A robust selection requires spectroscopic data with a highly complete spatial coverage to find over-densities of galaxies that appear to reside in a common structure.  Such a data set is provided by GAMA, a highly complete spectroscopic survey of nearly 300,000 galaxies down to magnitude $r < 19.8$. The full survey covers a total area of about $286$ degrees$^2$ split into five different patches on the sky \citep{Driver2009,Driver2011,Liske2015}. We restrict our analysis to the three equatorial fields of the survey, G09, G12 and G15 (which comprise a total area of 180 degrees$^2$), because there is a uniform target selection in these fields. Moreover, these fields overlap with the imaging data from KiDS \citep{deJong2017}, which are used to determine the weak lensing mass estimates of the groups.

The groups are identified using a Friends-of-Friends (FoF) algorithm in which galaxies are grouped based on their line-of-sight and projected physical separations \citep{Robotham2011}.  Unlike the standard GAMA group catalogue we applied a FoF algorithm on a approximately volume limited sample, we further discuss the FoF catalogue in Section~\ref{sec:FoF}. The integrated stellar masses of the group members, derived analogues to the integrated luminosity in \citet{Robotham2011}, are then used to select samples of groups for which we determine the ensemble averaged X-ray luminosity, tSZ signal and weak lensing mass.  We use weak gravitational lensing to determine the average group masses, because the amplitude of the lensing signal is proportional to the group mass. This signal itself is determined by measuring the coherent distortions in the shapes of galaxies in the background of the group \citep{Viola2015a}. This will be discussed in more details in Section 4.1.

\subsection{The BAHAMAS simulations}

To interpret the observations we rely on cosmological hydrodynamical simulations that can capture the complex baryon physics that determines the observed properties of galaxy groups. This requires a sufficiently large simulation volume to ensure a significant sample of massive halos that can be studied, but also sufficiently high resolution to study the small scales where baryonic processes are important.  Although the dynamic range of such simulations is rapidly increasing, current cosmological simulations must implement subgrid prescriptions to capture important physical processes that occur on scales that are too small to resolve directly (e.g., star formation, accretion onto black holes, initiation of feedback-driven outflows, etc.).  The OverWhelmingly Large Simulations (OWLS) project \citep{Schaye2010}, and its large-volume extension cosmo-OWLS \citep{LeBrun2014,McCarthy2014a}, highlighted the sensitivity of the predicted properties of collapsed structures to the details of the subgrid modelling. On large scales and for the massive haloes of interest here, this sensitivity is tied mostly to the modelling of AGN feedback as opposed to that of stellar feedback, which is dominant in lower mass systems \citep{McCarthy2011,LeBrun2014,Crain2015}.

The lack of ab initio predictive power of cosmological simulations when it comes to the stellar fractions of haloes led \citet{Schaye2015} to the conclusion that the feedback should be calibrated to reproduce these observations, motivating the Evolution and Assembly of GaLaxies and their Environment (EAGLE) project, a successor to OWLS.  In this approach, one can then run different models that are all calibrated on the same observables and test their realism by looking at other independent observables \citep{Crain2015}.  More recently, this calibration philosophy has been applied to larger scales in the BAHAMAS project \citep{McCarthy2017}.  \citet{McCarthy2017} which extended the calibration to also include the gas fractions of groups and clusters, since the hot gas dominates over the stellar mass fraction in such systems and is therefore crucially important when trying to constrain feedback models.

BAHAMAS consists of a suite of large-volume (400 Mpc/$h$ on a side cube) simulations with 1024$^3$ baryon and CDM particles and a force softening of 4 kpc/$h$, run in a variety of background cosmologies while adopting a fixed calibrated feedback model. 
Here we use the WMAP9-based cosmology run (with massless neutrinos), described in \citet{McCarthy2017}.  BAHAMAS was run using a modified version of the {\sc Gadget 3} code \citep{Springel2005}.  The simulations include subgrid treatments for metal-dependent radiative cooling \citep{Wiersma2009a}, star formation \citep{Schaye2008}, stellar evolution and chemodynamics \citep{Wiersma2009b}, and stellar and AGN feedback \citep{DallaVecchia2008,Booth2009}, developed as part of the OWLS project (see \citealt{Schaye2010} and references therein).  
The large volume of BAHAMAS means that the simulations contain the full range of massive haloes ($10^{12}-10^{15} M_\odot$), ideal for our purpose.  Importantly, \citet{McCarthy2017} have shown that BAHAMAS approximately reproduces the stacked baryon scaling relations found for the LBG sample by \citet{Planck2013a} for the tSZ effect and by \citet{Anderson2015a} for the X-ray luminosity. Our paper presents the next step, comparing the scaling relations of a galaxy group sample and comparing these to observations. 

Light cones of $5\times5$ deg$^2$ of the gas, stellar, and dark matter particles, along with the corresponding galaxy and halo catalogues, are constructed by stacking randomly rotated and translated simulation snapshots along the line of sight between $z=0$ and $z=3$ \citep{McCarthy2014a}.  We use 25 quasi-independent light cones constituting a total sky area of 625 $\mathrm{deg^2}$.  Figure \ref{Simsnapshots} shows cut outs of the tSZ-, X-ray- and lensing convergence maps ($\kappa$-maps) of one of the light cones, centred on the most massive cluster in one of the light cones ($\log[M_{500}/(h^{-1}M_\odot)]=14.3$). The Compton-$y$ signal is a direct integral of the gas pressure along the line of sight, whereas the X-ray map is produced by computing the X-ray spectrum for each gas particle in the simulation based on the gas pressure, temperature and metallicity before doing the line of sight integral. The $\kappa$-map, which is proportional to the projected mass, is computed using the KiDS source galaxy redshift distribution $n(z)$ \citep{Hildebrandt2016a}. 

\subsubsection{Galaxy selection prior to group finding}

\begin{figure}
\includegraphics[width=8.7cm]{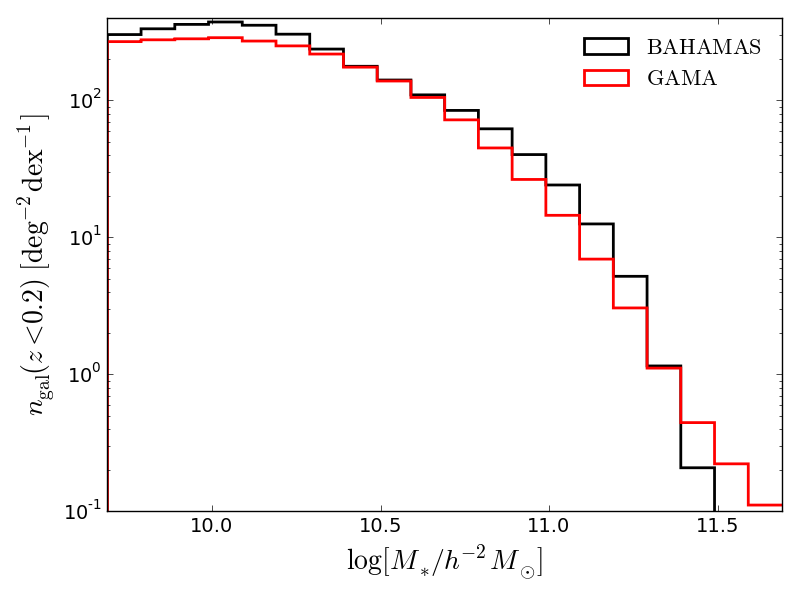}
\caption{The stellar mass functions of the approximately volume limited galaxy samples from GAMA and BAHAMAS. The simulation feedback models were calibrated to reproduce the present-day stellar mass function in \citet{McCarthy2017}.}\label{GSMF}
\end{figure}

Since we are attempting to compare the observed and predicted properties of optically-selected groups, the simulations should at least broadly reproduce the properties of the galaxy population (specifically the stellar masses). Otherwise we would likely select systems of different halo mass in the simulations and observations (and would therefore have no right to expect similarity in the gas-phase properties).  As noted above, the feedback model in BAHAMAS was calibrated to approximately reproduce the local galaxy stellar mass function (GSMF) as determined using SDSS data.  Here we compare to the GSMF from GAMA \citep{Wright2017}, where we select a sample of galaxies with a Petrosian stellar mass \citep{Taylor2011} $\log(M_*/M_\odot)>10$ and also implement a redshift cut of $z<0.2$.  The stellar mass limit is set by the resolution of the BAHAMAS simulations, while the redshift limit corresponds approximately to the maximum redshift out to which a passive galaxy of this mass can be observed given the depth of GAMA.  In other words, this selection corresponds to an approximately volume-limited sample\footnote{In fact, the maximum redshift out to which a $\log(M_*/M_\odot)=10$ galaxy will be included in GAMA is closer to $z\approx0.155$.  We have extended the sample to include all systems out to $z=0.2$, so the sample is not strictly volume-limited.  However, our results and conclusions do not change significantly when adopting the lower redshift cut, so we use the full sample and refer to it as an approximately volume-limited sample.} for galaxies of $\log(M_*/M_\odot)=10$.  In the simulations, the stellar mass is measured within a simple 30 kpc radius in 3D space, which both \citet{Schaye2015} and \citet{McCarthy2017} have found approximates the Petrosian stellar mass estimate well.  The corresponding GSMF from BAHAMAS agrees rather well with the observations (black histogram in Fig.~\ref{GSMF}), especially in comparison to many previous simulation efforts (see, e.g., the right panel of Fig.~5 in \citealt{Schaye2015}).

Note that in principle we do not have to restrict our analysis to a volume-limited sample with $z < 0.2$, but could instead use the full flux-limited sample of GAMA (modulo galaxies with masses below the resolution limit of BAHAMAS). This would allow us to probe groups and clusters at $z > 0.2$ and therefore boost our statistics. However, to mimic such a selection in the simulations requires the use of detailed stellar population models (which have non-negligible uncertainties) to calculate the flux of each galaxy, while also accounting for dust attenuation, K-corrections, etc. By restricting the analysis to an approximately volume-limited sample with a limited redshift range, our results are more robust against these modelling uncertainties.


\label{sec:GABA}

\section{Friends-of-Friends group selection}\label{sec:FoF}
Various algorithms have been used to find groups in observational data (e.g. \citealt{Eke2004a,Yang2005,Yang2007,Robotham2011}), as well as simulations (see \citealt{Knebe2011} and references therein).  Each approach has its strengths and drawbacks, assigning different weights to different quantities. Consequently,  comparisons between simulated and observed group samples are generally not trivial, unless the selection procedure used for the observations and simulations is the same.  However, as already noted, even adopting an identical group selection does not guarantee a useful comparison if the simulations do not have a broadly realistic galaxy population.  As a consequence of calibration on the observed GSMF, BAHAMAS does have a realistic galaxy population (see Section \ref{GAMA}), so we can now proceed to test the simulations against other independent data sets (namely the weak lensing masses, X-ray luminosities $L_X$, and tSZ $Y$, quantities which are defined later in Section 4.1, 4.2 and 4.3).

\subsection{Group selection set-up}\label{subsec:groupselect}

Here we take advantage of the FoF algorithm developed by \cite{Robotham2011} for the GAMA survey.  A crucial aspect of FoF algorithms in general is the choice of linking lengths.  For the flux-limited GAMA sample, \cite{Robotham2011} implemented projected and line-of-sight linking lengths that depend on galaxy luminosity, in the sense that the maximum allowed linking lengths increase with increasing luminosity.  Via comparison to mock galaxy catalogues, they found that having such a dependence (rather than having fixed linking lengths) yields a FoF sample that is closer in its statistical properties to the true underlying mock catalogue.

For the present comparison, we use an approximately volume-limited sample and adopt a fixed linking length, evaluated using eqns.~(1) and (4) of \citet{Robotham2011}.
We have found that BAHAMAS does not reproduce the GAMA galaxy r-band luminosity function, in that it predicts too many galaxies (by about a factor of 2) above the GAMA flux limit.  We have found that this is due to the simulated galaxies having too high star formation rates at late times (see McCarthy et al. 2017), which significantly boosts the optical luminosities (sometimes referred to as "frosting").  For example, star formation within the past 2 Gyr accounts for $<$ 5\% of the stellar mass in BAHAMAS galaxies but accounts for more than 25\% of the total r-band luminosity.  As the selection of the sample we study here does not depend on the luminosity function, this does not affect our conclusions, but it does prevent the use of a luminosity-dependent linking-length.  However, it is possible to adopt a stellar mass-dependent linking length, which we plan to do in future studies.  It also worth pointing out that for previous galaxy mocks used for GAMA, which were were derived from a semi-analytic model (see Robotham et al. 2011), the predicted luminosities were adjusted by hand to reproduce the observed luminosity function, whereas we have not attempted to do this for BAHAMAS.

Note that we use the observed GSMF with a lower stellar mass limit of $10^{10}$ M$_\odot$ to evaluate the mean comoving intergalaxy separation in eqn.~(2) of that study. Because we are adopting a different linking strategy than that of \cite{Robotham2011}, the group catalogue derived from the GAMA data were recomputed, so as to ensure a consistent comparison with the BAHAMAS group catalogues.  We also note that when computing the mean intergalaxy separation required for the linking length calculation, we use the same GSMF (the observed one) for both the data and the simulations.

We follow \cite{Robotham2011} and compute the comoving transverse and radial linking lengths as b$\langle {\rm L_{sep}} \rangle $ and b R$\langle {\rm L_{sep}} \rangle $ , respectively.  Here $\langle {\rm L_{sep}} \rangle $  is the mean comoving separation of galaxies with M$_\star ~ >$ $10^{10}$ M$_\odot$ and $z < 0.2$ (i.e., within our approximately volume-limited sample).  We compute $\langle {\rm L_{sep}} \rangle = 4.1 Mpc/h$ (note that this is mainly determined by the lowest-mass galaxies in our sample, as they are the most abundant). The coefficents b and R are adjustable.  Using mock galaxy catalogs generated using a semi-analytic model of galaxy formation, Robotham et al. deteremined the optimal values of b and R, by comparing the recovered group catalog (and its galaxy membership) to the true group catalog from the simulations.  Following \cite{Robotham2011}, we also adopt b=0.06 and R=18 for our analysis.  In principle we could perform similar tests to those in \cite{Robotham2011} to determine the optimal values of b and R for BAHAMAS, but leave this for future work.  While the values we adopt may not be optimal, they are consistently applied to both the simulated and observed galaxy catalogs, making the comparison a fair one. Thus, we use {\it exactly the same linking lengths} when deriving the FoF/group catalogues for BAHAMAS and GAMA, allowing us to make a fair and meaningful comparison between the two.  


\begin{figure*}
    \centering
    \subfloat[][Galaxy Stellar Mass Function]{
        \includegraphics[width=8.7cm]{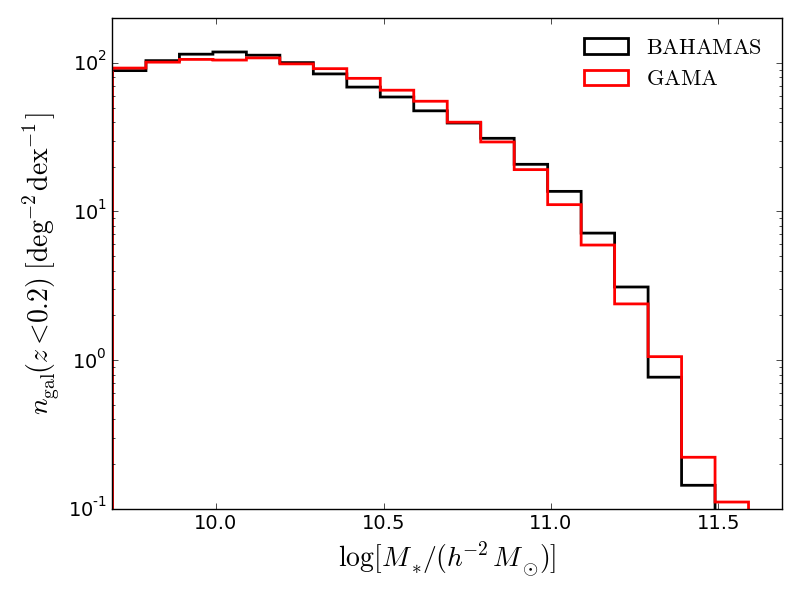}
    }
	~
    \subfloat[][Groups Stellar Mass Function]{
        \includegraphics[width=8.7cm, trim=-0.2cm 0 0.4cm 0]{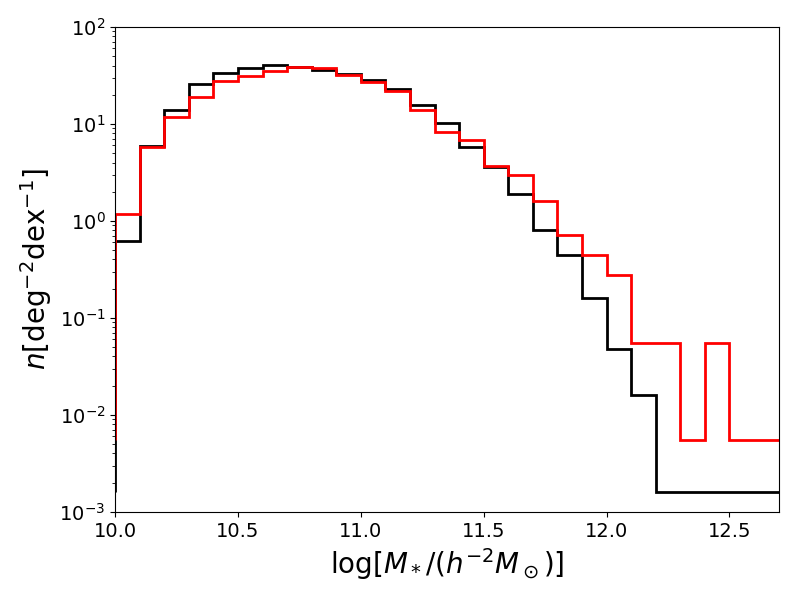}
    }
	    
    \subfloat[][Distribution of Projected Radii]{
        \includegraphics[width=8.9cm, trim=0 0.2cm 0 0]{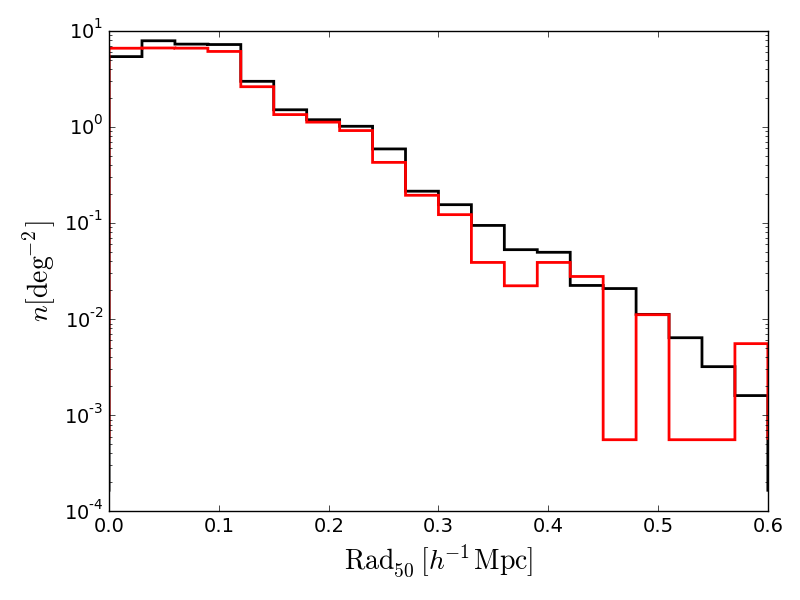}
    }
    ~ 
    \subfloat[][Groups Richness]{
        \includegraphics[width=8.7cm]{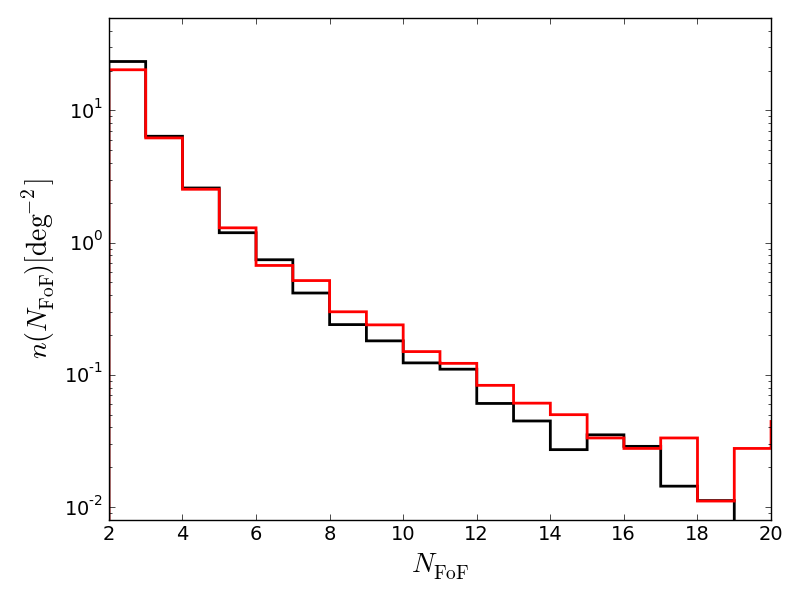}
	}
    \caption{Comparison of the group properties of the GAMA (red) and BAHAMAS (black) groups and galaxies. {\it Top left} shows the Galaxy Stellar Mass Function (GSMF) for the galaxies as associated with a Friends-of-Friends (FoF) group. {\it Top right} shows the integrated stellar mass function of the FoF groups. {\it Bottom left} shows distribution of projected radii ($\mathrm{Rad_{50}}$ as defined in \citet{Robotham2011}) of the identified FoF groups and the {\it bottom right} shows the multiplicity (`richness') function of the groups.}\label{fig:DataSims}
\end{figure*}

In Figure~\ref{fig:DataSims} we compare various distribution functions of the  GAMA (red) and BAHAMAS (black) groups using the modified FoF algorithm.  The top left panel shows the GSMFs of galaxies that are associated with FoF groups.  A comparison to Fig.~\ref{GSMF} indicates that approximately one-third of the galaxies with stellar masses of $> 10^{10}$ M$_\odot$ and $z < 0.2$ are assigned to groups by the FoF algorithm. The top right panel shows the total integrated group stellar mass function, where the total integrated group stellar mass is the summed stellar mass of all member galaxies, corrected for the GAMA luminosity function to account for missing flux of galaxies, analogues to the group $r$-band luminosity defined in \citet[section 4.4; eq. 22]{Robotham2011}.
We should note that, in order to have a perfect match with the data, and, assuming the cosmology is correct, it is necessary that the simulations fulfill the following conditions:

i) the simulated cosmology should be correct, as this sets the abundance of host dark matter haloes

ii) the simulated stellar mass-halo mass relation should be correct for the full galaxy sample, as this is required to select galaxies in the same way in GAMA and BAHAMAS

iii) the simulated distribution and abundance of satellites in groups and clusters of galaxies should be correct, otherwise the richness would be incorrect

iv) that both GAMA and BAHAMAS probe the same large-scale environments (characterised by e.g. the mean mass density).

Given that only (ii) was calibrated carefully in BAHAMAS, it is remarkable that all items (i) to (iv)  align with the data without forced calibration. Figure~\ref{fig:DataSims} indicates that the FoF group selection yields galaxy groups with similar properties in the data and simulations. The remaining differences are unlikely to affect the conclusion of this paper because we are looking at the *internal* properties of haloes (tSZ, X-ray, weak lensing mass) and not the abundance of haloes.

\subsection{The final group samples}\label{subsec:groupsamples}

\begin{table*}
\centering
\caption{Mean properties GAMA groups divided into three bins of based on their total group stellar mass. This table presents the bin edges and main statistics. The first column gives the bin limits, the number of groups is shown in the second column, the third and fourth column give the median redshift of the groups and the mean stellar mass of the BCG. The latter property is used in the modelling of the gravitational lensing signal. \label{table:binningschemes_GAMA}}

\begin{tabular}{c|rcc|rcc}
\hline
&&GAMA&&&BAHAMAS&\\
\hline
$\log[\frac{M_*^\mathrm{grp}}{h^{-2}\mathrm{M_\odot}}]$ & $N_\mathrm{groups}$ & $\bar{z}$ & $\log(\langle \frac{M_*^\mathrm{BCG}}{h^{-2} M_\odot}\rangle)$ 
& $N_\mathrm{groups}$ & $\bar{z}$ & $\log(\langle \frac{M_*^\mathrm{BCG}}{h^{-2} M_\odot}\rangle)$\\
\hline
 10.5 - 11.5	 & 518 & 0.147 & 10.77 & 1658 & 0.148 & 10.65\\
 11.5 - 11.8 & 137 & 0.167 & 11.03 & 347  & 0.161 & 10.99\\
 11.8 - 12.7   & 30  & 0.165 & 11.12 & 44   & 0.170 & 11.03\\

\hline
\end{tabular}

\end{table*}

For our stacking analysis, we divide the groups into 3 bins of total group stellar mass, $M_*^\mathrm{grp}$, which is defined as the sum of the stellar masses of the member galaxies, corrected for missing flux based on the GAMA luminosity function \citep[section 4.4; eq. 22]{Robotham2011}. We then measure the lensing signal for each bin and compute $M_{200}$ and $M_{500}$ from a halo model MCMC fit. With the halo masses and corresponding radii defined, we can measure the stacked tSZ and X-ray signal in each of the $M_*^\mathrm{grp}$ bins and combine them with the average halo masses from the lensing measurement to obtain the $Y-M_{500}$ and $L_X-M_{500}$ relations, where $Y$ is the integrated tSZ signal and $L_X$ is the X-ray luminosity (see Section 4.2 and 4.3 for the exact definition).

Table \ref{table:binningschemes_GAMA} provides details of each of the stellar mass bins. As commonly found in the literature (e.g. \citet{vanUitert2017}), we adopt the BCG as the operational definition of the group centre. Furthermore, \citet{Robotham2011} and \citet{Viola2015a} who found this is a good assumption using mock catalogues and weak lensing measurements, respectively. In Section \ref{Results} we discuss the validity of this assumption. 


\section{The stacked properties of the galaxy groups}\label{sec:DataStacking}
Having identified the galaxy groups, we proceed to measure their mean halo masses and diffuse gas content. The halo masses are determined using weak gravitational lensing, and the analysis is described in section~\ref{sec:Lensing}.
We explore two probes of the diffuse gas of the  intragroup medium, namely the X-ray emission (section~\ref{sec:Xray}) and the tSZ effect (section~\ref{sec:tSZ}).
The two tracers have differing dependencies on gas density, temperature, and metallicity, making them complementary probes.  


\begin{figure*}
\centering
\includegraphics[width=1.\textwidth,trim=+4cm 0 4cm 1cm]{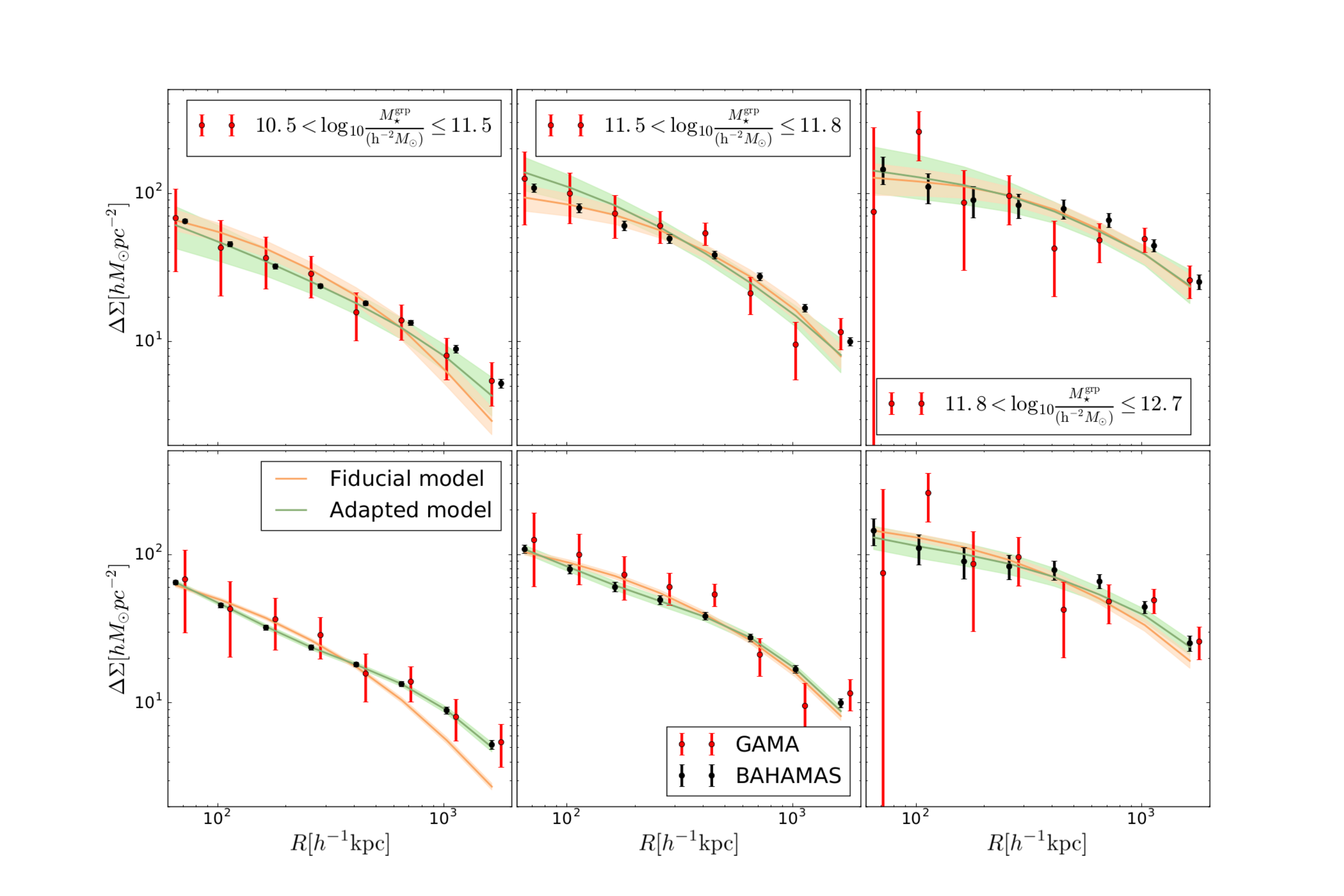}
\caption{The stacked excess surface density (ESD) profile of the FoF groups for the three stellar mass bins. The red points correspond to measurements around the GAMA groups using the KiDS weak lensing data while the black points are the signal as measured from the FoF groups found in the BAHAMAS simulation. In the {\it top panels} the fits to the actual data are presented, and the {\it bottom panels} show fits to the simulations. The halo model fits and their 68\% confidence regions are indicated by the coloured regions. The fiducial or standard halo model is indicated by orange regions, whereas the results from the modified or adapted model are shown in green.
The similarity between the ESD profiles of the GAMA and BAHAMAS groups is remarkable and further highlights the level of realism of the simulations as well as our ability to select the same objects.  } \label{fig:ESDcomparison}
\end{figure*}
\subsection{Weak gravitational lensing}\label{sec:Lensing}

The images of distant galaxies are distorted by the tidal effect of the gravitational potential of intervening matter; this effect is commonly referred to as weak gravitational lensing \citep[see e.g.][for a detailed introduction]{Bartelmann2001}, and has become a widely used tool to study the matter distribution in the Universe. The amplitude of the signal is directly related to the mass of the lens, irrespective of its dynamical state. This makes it an ideal technique to determine the masses of massive objects such as galaxy groups. Unfortunately, the lensing signal for individual groups is too weak to be detected and therefore we can only study ensemble averages.

\cite{Viola2015a} studied the lensing signal of GAMA groups and we refer the interested reader there for a more detailed discussion of the measurements and modelling therein. The amplitude of the group signal at a projected distance $R$ from the group centre is directly related to the excess surface density, $\Delta\Sigma$, defined as:
\begin{equation}
\Delta\Sigma(R)=\bar\Sigma(\le R)-\bar\Sigma(R),
\end{equation}
where $\bar\Sigma(\le R)$ is the mean surface density within an aperture of radius $R$, and $\bar\Sigma(R)$ the azimuthally averaged surface density at radius $R$. The excess surface density can be expressed in terms of the 
azimuthally averaged tangential shear, $\gamma_\mathrm{t}$, and the critical surface density, $\Sigma_{\rm crit}$:
\begin{equation}
\Delta\Sigma(R)=\Sigma_{\rm crit}\langle\gamma_\mathrm{t}\rangle(R),
\end{equation}
where the inverse critical surface density for a lens at redshift $z_l$,
and sources with a redshift distribution $n(z_\mathrm{s})$, is given by
\begin{equation}
\Sigma^{-1}_{\mathrm{crit}} =\frac{4\pi G}{c^2}\int_{z_\mathrm{l}+0.2}^{\infty} \frac{D_\mathrm{l}(z_\mathrm{l})D_\mathrm{ls}(z_\mathrm{l},z_\mathrm{s})}{D_\mathrm{s}(z_\mathrm{s})}n(z_\mathrm{s})\mathrm{d}z_\mathrm{s},
\label{eq:crit}
\end{equation}
where $D(z)$ is the angular diameter distance to redshift $z$, and $G,c$ are respectively the gravitational constant and the speed of light. We only considered sources with a redshift $>z_\mathrm{l}+0.2$; this mitigates 
any effects which might be caused by the contamination of the source galaxy sample by group members \citep{Dvornik2017}.

The tangential shear is obtained using the shape measurements from the KiDS $r$-band data \citep{Hildebrandt2016a,deJong2017}. The shapes themselves were determined using the well-characterized {\sc lens}fit 
algorithm \citep{Miller2007, Conti2017} and the residual systematic error on statistical shear measurements is about 1\%.
We follow \cite{Viola2015a} to compute the signal from the data and correct the signal for residual multiplicative bias. For the lenses we use the 
group spectroscopic redshift $z_\mathrm{l}$ as measured by GAMA. The source redshift distribution, $n(z_\mathrm{s})$, is determined by directly calibrating the KiDS photometric redshifts using deep spectroscopic data; we refer to Section 3.2 of \citep{Hildebrandt2016a} for further details. 

We compute the simulated shear maps for each BAHAMAS light cone, adopting the KiDS source redshift distribution. This enables us to compare the results directly to the observations. Figure \ref{fig:ESDcomparison}
shows the resulting stacked Excess Surface Density (ESD) profiles for the three stellar mass bins. The red points with error bars indicate the actual measurements of the excess surface density around GAMA groups. The black points with (small) error bars correspond to the signal measured from BAHAMAS.  The error budget of the GAMA data is computed using the analytical approach described in \citet{Viola2015a}, whereas the simulation errorbars are found using a bootstrap resampling. As the latter only captures the variance within a bin, since shape noise is absent in the simulated shear maps we use, the errorbars on the BAHAMAS measurement are much smaller. 
The similarity in the lensing signal of the GAMA and BAHAMAS groups demonstrates our ability to select groups consistently between observations and hydrodynamical simulations.

Following \cite{Viola2015a}, we use a halo model \citep[e.g.][]{Seljak2000,Cooray2002} to interpret the stacked excess density profiles $\Delta\Sigma(R)$ presented in Fig.~\ref{fig:ESDcomparison}. In doing so, we assume that each galaxy group resides in a dark matter halo and that the stacked $\Delta\Sigma(R)$ profile can be modelled using a statistical description of how galaxies are distributed over dark matter haloes of different mass, and how these haloes cluster. We use a Navarro Frenk White profile \citep{NFW1996} to describe the density  profile of dark matter halos, adopting the \cite{Duffy2008a} mass-concentration relation. We describe the halo occupation distribution (HOD hereafter) of galaxy groups as a log-normal distribution in mass. We include in the modelling a mis-centring term to account for a possible displacement of the BCG, which is used as a proxy for the group centre from the bottom of the group's potential well. Finally, we describe the clustering of the halos using the halo mass function and the halo bias function from \citet{Tinker2010}. We refer the reader to  \cite{Viola2015a} for a more detailed description of our implementation.

In the standard version of our halo model we jointly fit the $\Delta\Sigma(R)$ profiles in the three stellar mass bins. The free parameters are the amplitude of the NFW mass-concentration relation (1 parameter), the width (1 parameter) of the log-normal HOD and its mean in each of the three bins (3 parameters), the probability for the BCG to be mis-centred (1 parameter) and the amount of mis-centring from the bottom of the gravitational potential well (1 parameter). The priors we used for those parameters are the same as in \cite{Viola2015a}. 

We fit this halo model to both the BAHAMAS and the real data and show the best-fit model and the 68 percent confidence intervals in Figure~\ref{fig:ESDcomparison} in orange. The top panels show the results when the model is fit to the GAMA measurements, whereas the fits to BAHAMAS are shown in the bottom panels. Given the small error bars on the BAHAMAS signal, it is apparent that the model is a poor description of the signal. In particular, the model fails to describe the signal in the first stellar mass bin where the effect of the fragmentation/aggregation of true halos caused by the FoF algorithm (see appendix \ref{FoFvsTrue}) is worst. Moreover, we retrieve  halo masses that are biased high by 0.1-0.3 dex depending on the stellar mass bin.

We therefore also explore a an adapted version of the halo model in which each stellar mass bins is fitted independently and a larger prior for the amount of mis-centring is employed. This version of the halo model has eight more parameters than the standard one and hence it has significantly more freedom in fitting the signal. We fit this model to both the  BAHAMAS and the real measurements and we show the best-fit models and the 68 percent confidence intervals in green in Fig.~\ref{fig:ESDcomparison}.
As before, the top panel shows the results of the fit to the signal around GAMA groups and the bottom panels show the fits to the simulated data. Nevertheless, in this case we find that halo masses are nearly unbiased in the three stellar mass bins, although this result comes at the expense of precision (errors on the masses are larger by a factor of two). It is important to keep in mind that this extended model is designed to provide a good fit to the data and self-consistent masses despite the fragmentation problem (see appendix \ref{FoFvsTrue}). Consequently, its parameters do not provide physical insight in to the mass structure of the groups. In the rest of the paper we use the adapted halo model unless otherwise specified. In section 6, where we investigate the effects of fragmentation/aggregation on the scaling relations, we compare the performance of both halo models against the true simulation data.

Finally we list the halo masses obtained from both HOD models in tables \ref{table:results_GAMA} and \ref{table:results_BAHAMAS} for GAMA and BAHAMAS respectively, and appendix \ref{HaloModelFits} lists all fitted parameters to the simple and adapted halo models).

\subsection{X-ray emission from hot gas in galaxy groups\label{sec:Xray}}

Within the potential well of the galaxy group haloes, thermal Bremsstrahlung, in case of the most massive ones ($T_\mathrm{gas} \sim 10^8 \mathrm{K}$), and metal-line emission ($T_\mathrm{gas} \lesssim 10^7 \mathrm{K}$) provide effective mechanisms for gas to radiate away some of its thermal energy (\citealt{Bertone2010,VandeVoort2013} and references therein). We study the resulting X-ray luminosities of the groups using data from the ROSAT All-Sky Survey (RASS) \citep{Voges1992}. RASS is an all-sky survey in the soft band X-ray survey conducted with the position sensitive proportional counter instrument (PSPC) aboard the Röntgensatellit (ROSAT) \citep{Truemper1986, Truemper1992}. In this work we use a full sky map of the ROSAT data, made publicly available by the Centre d'Analyse de Données Etendues (CADE)\footnote{See: \url{http://cade.irap.omp.eu/dokuwiki/doku.php?id=welcome}}. These maps are provided in the \texttt{HEALPix} pixelisation scheme \citep{Gorski2005}.
    
CADE provides RASS photon count maps in three energy bands as well as a map of the exposure time. The three photon count maps cover (1) the full ROSAT energy range of 0.1 - 2.4 keV, (2) the softest X-ray radiation in the range of 0.1 - 0.4 keV and (3) the 0.5 - 2.4 keV energy band. In this study we use the latter band, as below 0.5 keV photons suffer heavily from absorption by the interstellar medium of the Milky Way. We measure the stacked X-ray luminosities of both the GAMA and BAHAMAS galaxy groups by performing an aperture photometry procedure similar to the method outlined in \citet{Anderson2015a}. 

For each group we measure the X-ray flux in an aperture centred on the BCG. We start the extraction of the signal by estimating the halo mass, $M_{200}$, of the group based on its integrated group stellar mass and the $M_{200} - M_{*,\mathrm{grp}}$ relation we obtained from the weak gravitational lensing measurement (see Section~\ref{sec:Lensing}). Here we have defined $M_{200} \equiv 200 \times 4/3\pi R_{200}^3\rho\crit(z)$, where $\rho\crit(z)$ is the critical density of the Universe at redshift $z$. We then calculate $M_{500}$ and $R_{500}$,(with $M_{500}$ and $R_{500}$ defined analogous to $M_{200}$ and $R_200$), assuming an NFW density profile and the best-fit effective concentration parameter, $c_\mathrm{m}^\mathrm{eff}$, from the best fit halo model (see \citealt{Viola2015a} for details). With the radius $R_{500}$ defined, we then extract a circular aperture around the group's position from the X-ray map with angular radius $\theta_\mathrm{extract} = 2 \theta_{500}(R_{500},z) + \mathrm{FWHM_{RASS}}$, where $\theta_{500}(R_{500},z) =  R_{500}/d_A(z)$, with $d_A(z)$ the angular diameter distance to redshift $z$ and $\mathrm{FWHM_{RASS}}$ is the $1.8'$ full width half maximum of the RASS. The group signal is then computed as the sum of the photon counts within $\theta_{500}$ minus the local background defined an annulus between $1.5 \times \theta_{500}$ and $\theta_\mathrm{extract}$. 
	
Note that we do not apply a point spread function (PSF) correction to the measured luminosity since \citet[Fig.~4]{Anderson2015a} have shown that the PSF of ROSAT is more compact than $\theta_{500}$ for the mass range of the systems we study here (which are all at $z < 0.2$). The PSF will therefore have a negligible effect on the total flux within the aperture $\theta_{500}$. We do not mask bright sources, because we show in Appendix \ref{Xpointsources} that their contribution is not significant within our current uncertainties.
				
Having measured the background subtracted photon count-rates for each group, we convert these into a physical rest-frame flux using the web tool webPIMMS provided by NASA's High Energy Astrophysics Science Archive Research Center (HEASRAC) \footnote{https://heasarc.gsfc.nasa.gov/cgi-bin/Tools/w3pimms/w3pimms.pl}. The conversion factors are provided in Table \ref{table:results_GAMA}. Details on the conversion of photon counts to flux can be found in Appendix \ref{sec:counts2flux}. Finally, we stack the resulting luminosities of the groups in the stellar mass bins and estimate the error by employing a bootstrap re-sampling over the sample in the bin. This uncertainty on the signal captures the statistical error on the mean as well as the sampling variance of the sample within the bin, the latter of which is the dominant source of uncertainty (we ignore cosmic variance in this study). Finally, to test our stacking analysis against possible systematic errors, we conducted a null-test by stacking random positions, details of which are provided in Appendix \ref{NullTest}. We find our stacking procedure to be free of significant biases.

The aperture photometry procedure applied to the simulation data is virtually identical to the procedure outlined above, differing only in that the simulation X-ray maps are given in observer frame flux (instead of photon counts) and we therefore only $k$-correct these into the rest frame flux (see Appendix \ref{sec:counts2flux} for details). The $k$-corrections are given in Table \ref{table:results_BAHAMAS}. We note that we smooth, with a Gaussian kernel, the higher resolution simulation maps to the RASS resolution of $1.8'$. 

\subsection{The thermal Sunyaev-Zel'dovich effect in galaxy groups\label{sec:tSZ}}	

Thermal X-ray emission of the diffuse intragroup gas is highly sensitive to the gas density, it therefore is an excellent probe of the inner regions of groups and clusters, but a less efficient tracer of the outskirts.  The tSZ effect, which is a measure of the inverse Compton scattering of the low energy cosmic microwave background (CMB) photons by the highly energetic electrons of the intracluster medium, is, on the other hand, a more sensitive probe of the outskirts.  This is due to its weaker dependence on gas density (which is linear, rather than scaling as the square of the mass density as in the case of X-ray emission).  In this scattering process, a CMB photon gets an effective energy boost, changing its frequency which can be observed as a local distortion of the CMB spectrum \citep{Sunyaev1972}.
		
A common estimate of the tSZ effect is the Compton-$y$ parameter (the mean energy change of a photon due to scattering when travelling through a medium), integrated over the solid angle of the galaxy (-cluster) halo, $\D\Omega = \D A /d_A^{2}(z)$:

\begin{align}
	Y_\mathrm{c}^\mathrm{cyl}(M,z) &= d_A^{-2}(z) \frac{\sigma_\mathrm{T}}{m_\mathrm{e}c^2} 2\pi \int_0^{R_\mathrm{c}(M_{500})} \D R R \int_{0}^\infty \D l P_\mathrm{e}(R,M,z)\ .
\end{align}
Here $\sigma_\mathrm{T}$ is the Thomson scattering cross section, $m_\mathrm{e}c^2$ is the electron rest mass energy and $P_\mathrm{e}(r,M,z)$ is the electron pressure at a distance $r$ from the centre of a halo of mass $M$ at redshift $z$. $R$ is the projected distance to the centre of the halo and we have integrated the Compton-$y$ parameter over a cone of radius $R_\mathrm{c}$ at the group location. As Compton-$y$ is dimensionless, $Y_\mathrm{c}^\mathrm{cyl}$ has units of area and is commonly expressed in square arcminutes. 
		
We measure the tSZ signal of the GAMA galaxy groups using the all-sky Compton-$y$ map from the Planck Collaboration \citep{Planck2015}. The map is based on the Planck full mission data and, like the RASS maps, is provided in the \texttt{HEALPix} pixelisation scheme \citep{Gorski2005}. The Planck Collaboration published two different maps\footnote{\url{https://wiki.cosmos.esa.int/planckpla2015/index.php/Specially_processed_maps\#2015_Compton_parameter_map}}, which are the result of different tSZ reconstruction algorithms from the CMB temperature maps. \citep{Planck2015}. In this work we make use of the \texttt{MILCA} map and apply both the point source and 40\% galactic foreground masks. 
	  
We stack the cylindrical integrated tSZ signal $Y_\mathrm{c}^\mathrm{cyl}$ of the groups in bins of total stellar mass (see Table \ref{table:binningschemes_GAMA}).  In accordance with previous studies, we choose a cylinder radius 
$R_\mathrm{c} = 5 \times R_{500}$ to account for the relatively low resolution of the Planck $y$-map of 9.66 arcminutes, which is larger than the radius $\theta_{500}$ for the majority of the systems we study.

\begin{figure*}
\centering

    \subfloat{
        \includegraphics[width=8.7cm]{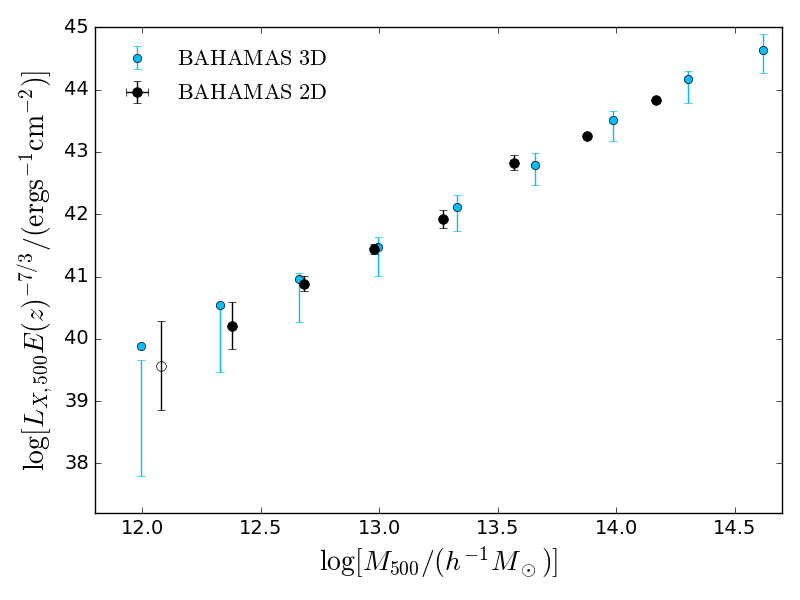}
    }
    ~
\subfloat{
        \includegraphics[width=8.7cm]{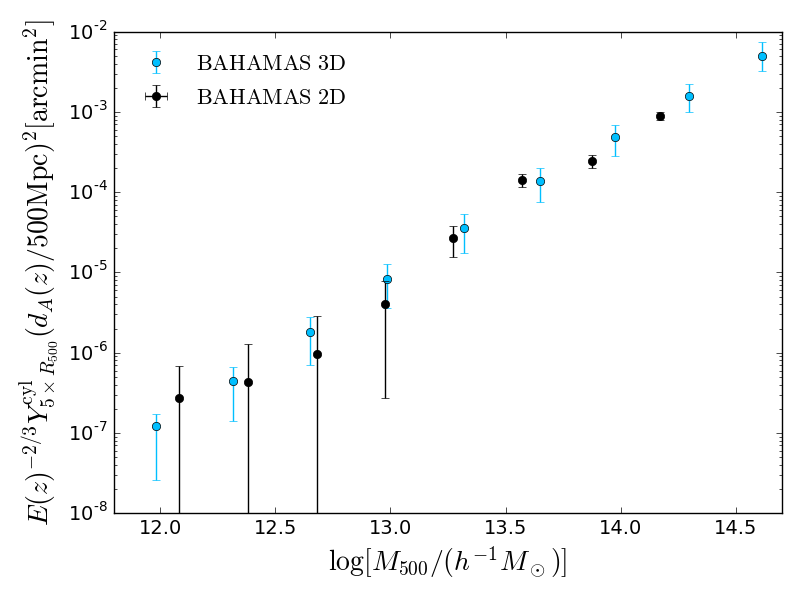}
    }
\caption{A comparison of the observationally processed simulation data and the projected 3D data from the BAHAMAS simulation. {\it Left:} The stacked soft band X-ray luminosity $L_{X,500}-M_{500}$ relation.  {\it Right:} The stacked $Y_{5\times R_{500}}-M_{500}$ relation.  The true (3D) mean relations are represented by the blue data points, while the observationally-processed (that is from the projected simulation) stacked relations are represented by the black data points.  The error bars of the black points (2D "observations") come from a bootstrap re-sampling analysis. The vertical lines on true (3D) points are not error bars in strict sense, but mark the 16th and 84th percentiles regions of the underlying sample. The open data point in the left panel show negative measured flux values.}\label{fig:2Dvs3D}
\end{figure*}

The stacking procedure we employ for the tSZ signal is very similar to the one used for the X-ray data.  Specifically, for every group in a stack we extract the pixels within an angular aperture $\theta_\mathrm{extract} = 5\times \theta_{500}(R_{500},z) + \mathrm{FWHM_{Planck tSZ}}$, where $\mathrm{FWHM_{Planck tSZ}}$ is the Planck Compton-$y$ map beam size. The tSZ signal is then measured inside an aperture $R_\mathrm{c}$ after subtracting the background estimated as the mean signal in an annulus between $R_\mathrm{c}$ and $R_\mathrm{extract}$. Next, the measured signal of the groups in the stellar mass bins is stacked and, analogous to the X-ray luminosity measurement, the error is calculated from a bootstrap re-sampling over the groups in a given bin. As with the X-ray stacking we tested our stacking procedure against possible systematics by conducting a null-test, the results of which are shown in Appendix \ref{NullTest}.

The measurement of the tSZ signal of the BAHAMAS galaxy groups is carried out analogously to the data. We use the maps constructed for each lightcone, which are smoothed with a Gaussian beam with a FWHM of $9.66'$ to match the beam size of the Planck Compton-$y$ map before applying the stacking procedure outlined above. 


\subsection{Testing the X-ray and tSZ stacking analyses}\label{sec:2Dvs3D}


We have tested our X-ray and tSZ stacking analyses for possible biases using the simulations, for which we can compute the true X-ray (3D) luminosity and tSZ signals and compare this to the stacked 2D analysis discussed above (which is applied to the observational data and the simulation light cone in an identical way).  Specifically, using the true groups in the simulations, we evaluate the mean X-ray luminosity-halo mass and tSZ-halo mass relations, using 3D spherical apertures of $r_{500}$ and $5 r_{500}$ for the X-ray and tSZ, respectively (see \citealt{McCarthy2017} for further details of how the tSZ and X-ray emission are calculated from the particles).  This 3D analysis is performed using a single snapshot of the simulation, output at $z=0.125$, without going via the light cones and aperture photometry procedure described in Sections 4.2 and 4.3.  For comparison, we then subject each true halo in the light cones to our 2D observational analysis and we compute the stacked X-ray luminosity$-$halo mass and tSZ$-$halo mass relations using the true halo mass.  The only difference between what we do here and what is described above in Sections 4.2 and 4.3 is that here we do not use an (observational) FoF algorithm to find the groups (we use the true simulation groups) and we use true halo masses rather than lensing masses.  This allows us to isolate any potential biases in our X-ray or tSZ stacking procedure (e.g. due to inaccurate background estimation, source confusion, etc.).

Figure \ref{fig:2Dvs3D} shows the comparison between the `observationally processed' or 2D data (black) and the simulation 3D data (blue) for the $L_X-M_{500}$ relation (left panel) and the $Y-M_{500}$ relation (right panel, where $\tilde Y$ is defined in Section 5). We note that, whereas the 3D data (blue) error bars show the 16-84 percentile region, the observationally processed data (black) shows the $1\sigma$ error bars from a bootstrap re-sampling.  

The true X-ray luminosity$-$mass relation in the left panel of Figure \ref{fig:2Dvs3D} is well recovered by our 2D observational analyses over the full range of (true) halo masses considered here. We note the fact that the mean 3D X-ray luminosities at low mass lie outside the 16$^{th}$ - 84$^{th}$ percentile interval implies that the signal in these bins is dominated by a small fraction of the systems with higher than typical luminosities.  
The $Y^\mathrm{cyl}-M_{500}$ relation from the simulated observations (black points) is statistically consistent with the true answer (cyan points) .

\begin{table*}
\caption{The mean group properties of the GAMA Friends-of-Friends groups from the stacking analysis. The first column gives the mean group stellar mass with the standard error on the mean, the second and third columns provide the best-fit $M^\mathrm{x}_{500}$ with the 16th and 84th percentile uncertainties based on MCMC simulation of the standard (S) and adapted (A) halo model respectively. The fourth column provides the counts to flux conversion factors $C_\mathrm{conversion}$ (note that this includes the $k$-correction) and the mean gas temperature (derived from $M^\mathrm{A}_{500}$) $\bar{kT}$ is given in the fifth column. Finally the stacked thermal Sunyaev-Zel'dovich signal and the X-ray luminosity with their $1\sigma$ uncertainties from the bootstrap analysis are provided in column six and seven.\label{table:results_GAMA}}
\begin{tabular}{cccccrc}\hline
$\log[\frac{M_*^\mathrm{grp}}{h^{-2}\mathrm{M_\odot}}]$ & $\log[\frac{M^\mathrm{S}_{500}}{h^{-1}M_\odot}]$ & $\log[\frac{M^\mathrm{A}_{500}}{h^{-1}M_\odot}]$ & $C_\mathrm{conversion}$ & $\bar{kT}$ & $Y^\mathrm{cyl}_{5\times R_{500}}$ & $\log[\frac{L_{X,500}}{\mathrm{erg\ s^{-1}}}]$ \\
($\mathrm{dex}$) & ($\mathrm{dex}$) & ($\mathrm{dex}$) & ($10^{-11} \mathrm{erg\ cm^{-2}cts^{-1}}$) & ($\mathrm{keV}$) & ($10^{-5} \mathrm{arcmin^{2}}$) & ($\mathrm{dex}$)\\ \hline
$11.23\pm0.01$ & $12.99_{-0.10}^{+0.09}$ & $13.29_{-0.17}^{+0.16}$ & $1.11$ & $1.02$ & $7.39\pm2.78$ & $41.60_{-0.33}^{+0.10}$ \\
$11.62\pm0.01$ & $13.46_{-0.07}^{+0.07}$ & $13.57_{-0.15}^{+0.21}$ & $1.22$ & $1.59$ & $17.72\pm8.31$ & $42.58_{-0.09}^{+0.19}$ \\
$11.94\pm0.03$ & $13.95_{-0.08}^{+0.08}$ & $13.92_{-0.11}^{+0.09}$ & $1.39$ & $2.76$ & $68.73\pm27.56$ & $43.32_{-0.08}^{+0.07}$ \\ \hline
\end{tabular}
\end{table*}

\begin{table*}
\caption{As Table \ref{table:results_GAMA} but for the BAHAMAS FoF groups.  As the simulation X-ray maps are provided in observed flux, only the $k$-correction is provided in column three. \label{table:results_BAHAMAS}}
\begin{tabular}{cccccrc}\hline
$\log[\frac{M_*^\mathrm{grp}}{h^{-2}\mathrm{M_\odot}}]$ & $\log[\frac{M^\mathrm{S}_{500}}{h^{-1}M_\odot}]$ & $\log[\frac{M^\mathrm{A}_{500}}{h^{-1}M_\odot}]$ & $k$ & $\bar{kT}$ & $Y^\mathrm{cyl}_{5\times R_{500}}$ & $\log[\frac{L_{X,500}}{\mathrm{erg\ s^{-1}}}]$ \\
($\mathrm{dex}$) & ($\mathrm{dex}$) & ($\mathrm{dex}$) &  & ($\mathrm{keV}$) & ($10^{-5} \mathrm{arcmin^{2}}$) & ($\mathrm{dex}$)\\ \hline
$11.21\pm0.00$ & $12.95_{-0.02}^{+0.02}$ & $13.35_{-0.03}^{+0.04}$ & $1.00$ & $1.13$ & $7.74\pm1.16$ & $42.40_{-0.14}^{+0.09}$ \\
$11.61\pm0.00$ & $13.47_{-0.03}^{+0.03}$ & $13.61_{-0.03}^{+0.04}$ & $0.96$ & $1.71$ & $16.55\pm3.50$ & $42.86_{-0.20}^{+0.10}$ \\
$11.89\pm0.01$ & $13.87_{-0.05}^{+0.06}$ & $13.93_{-0.07}^{+0.06}$ & $0.98$ & $2.78$ & $64.86\pm16.24$ & $43.74_{-0.07}^{+0.14}$ \\ \hline
\end{tabular}
\end{table*}




\section{Scaling relations}\label{Results}

In this section we present our main results, which are the recovered stacked scaling relations of GAMA and BAHAMAS groups.  We first present the scaling relations between the stacked signals (weak lensing mass, X-ray luminosity and tSZ flux) and the integrated group stellar mass (Figure ~\ref{fig:BaryonOnly}).  We then use the stacked weak lensing halo masses to derive the X-ray luminosity$-$halo mass and tSZ flux$-$halo mass relations (Figure ~\ref{fig:LxYM_adaptedMass}).

\subsection{Lensing, X-ray, and tSZ scalings with group integrated stellar mass}

\begin{figure}
\centering

\subfloat[][Halo mass - stellar mass relation for groups]{
        \includegraphics[width=8.5cm]{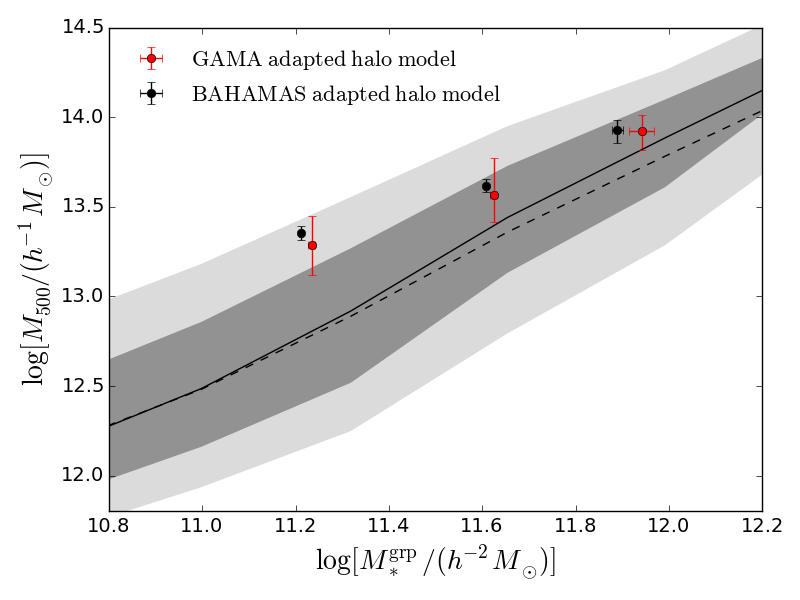}
    }
\\
\subfloat[][X-ray Luminosity - stellar mass relation for groups]{
        \includegraphics[width=8.5cm]{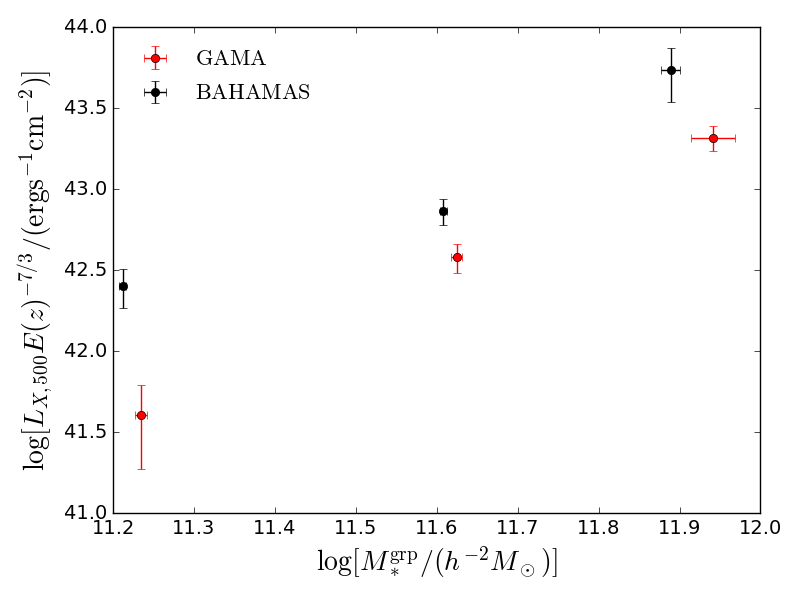}
        }
\\
\subfloat[][Projected y-Compton - stellar mass for groups]{
        \includegraphics[width=8.5cm]{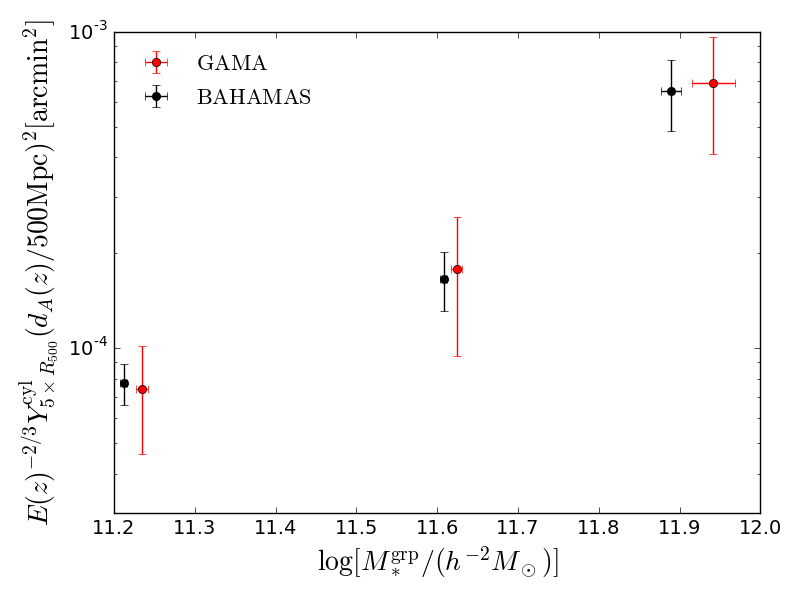}
    }
\caption{{\it Top panel:} A comparison of the stacked $M_{500} - M_\mathrm{*,grp}$ relation for the galaxy  groups of the GAMA survey (red) and the BAHAMAS simulations (blue). The grey area shows the 16-84 and 2.5-97.5 percentile regions of the true simulation data. The black solid line is the median (50th percentile) and the black dashed line shows the mean. {\it Middle panel:} Same as the top panel but for the soft band X-ray luminosity $L_{X,500}-M_\mathrm{*,grp}$. {\it Lower panel:} As above but for the $Y_{5\times R_{500}} - M_\mathrm{*,grp}$ relation. The agreement between the observations and simulation are excellent for the halo mass and tSZ scalings, but less so for the X-ray data.  We discuss a possible explanation in Section 5.2.}\label{fig:BaryonOnly}
\end{figure}

In the top panel of Figure \ref{fig:BaryonOnly} we show the stacked weak lensing mass ($M_{500}$) in bins of integrated group stellar mass.  Here we show the weak lensing masses derived using the more flexible adapted halo model of \citet{Viola2015a} described in Section 4.1. We find that the mean observed and predicted halo masses agree to better than 0.1 dex  for each of the three stellar mass samples. As noted previously, the error bars for the simulation data points are significantly smaller than those for the observational data because the simulated shear measurements neglect shape noise and the maps have a significantly higher source density.

The middle panel of Figure \ref{fig:BaryonOnly} shows the stacked X-ray luminosity as a function of total stellar mass. The $y$-axis includes a factor $E(z)^{-7/3}$ to scale out the effects of self-similar redshift evolution, where $E(z)=H(z)/H_0$ is the dimensionless Hubble parameter.  Here we find an amplitude offset between the measured and predicted X-ray luminosities at the level of $\sim 0.5\ \mathrm{dex}$.  We will discuss the possible origin of this discrepancy further below.

In the lower panel of Figure \ref{fig:BaryonOnly} we plot the relation between the average stellar mass of the groups
$M_*^\mathrm{grp}$ and $E(z)^{-2/3}Y_{5\times R_{500}}(d_A(z)/500\mathrm{Mpc})^2$, where the exponent of the dimensionless Hubble parameter assumes self-similarity. Note that we do not expect that the evolution of the X-ray or tSZ signals to be perfectly self-similar (e.g., \citealt{LeBrun2017,Barnes2017}).  We measure a clear stacked tSZ signal for both observations, signal to noise ratio (S/N) of 4.2, and simulations, S/N = 8.9, and find that the $Y-M_*$ relation between the simulation and observational data are statistically consistent.

We note that some previous studies (e.g., \citealt{Melin2010,Planck2013a,Sehgal2013}) present the tSZ signal as the Compton $y$ parameter integrated over a sphere of radius $R_{500}$, $Y_{500}^\mathrm{sph}$. However, the spherically-integrated Compton $y$ signal is not a directly observable quantity.  Converting the cylindrically-integrated $y$ signal to a spherically-integrated $y$ signal requires either a de-projection or some assumptions on the shape of the pressure profile of the gas. The former is not feasible with the resolution of the current data and it assumes that there is no line of sight contamination by foreground or background objects.  Previous studies, (e.g. \citealp{Melin2010, Planck2013a, Sehgal2013}), have adopted the so-called `universal pressure profile' (UPP) of \citet{Arnaud2010} in order to convert the observed signal into a spherically-integrated quantity.  However, the UPP, whilst providing a reasonably good description of very massive galaxy clusters, is not expected to describe the gas distribution in low-mass groups as well, due to the stronger impact of non-gravitational physics at this scale (e.g., \citealt{LeBrun2017}).  We therefore choose to present our results as $Y_\mathrm{c}^\mathrm{cyl}$, which is a directly observable property in both data and simulation.

\subsection{X-ray and tSZ scalings with lensing mass}

\begin{figure*}
\centering
    \subfloat{
        \includegraphics[width=8.7cm]{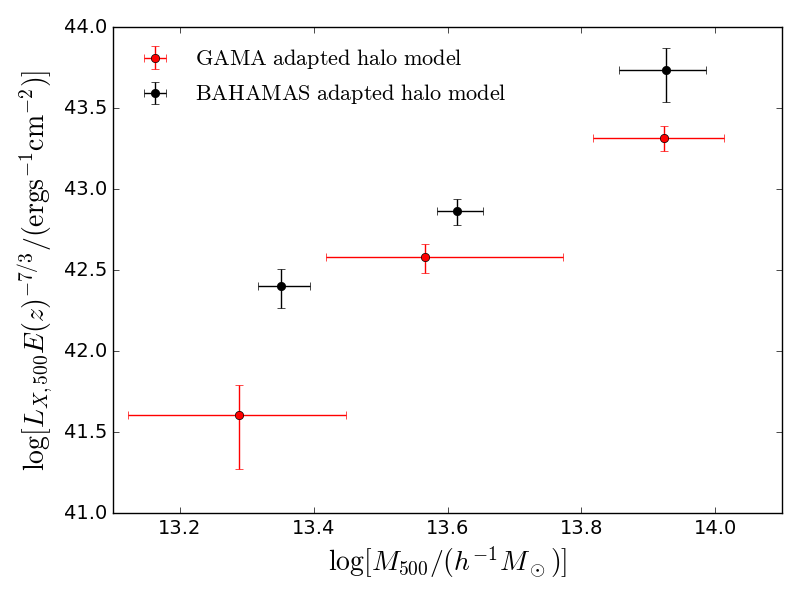}
    }
~
\subfloat{
        \includegraphics[width=8.7cm]{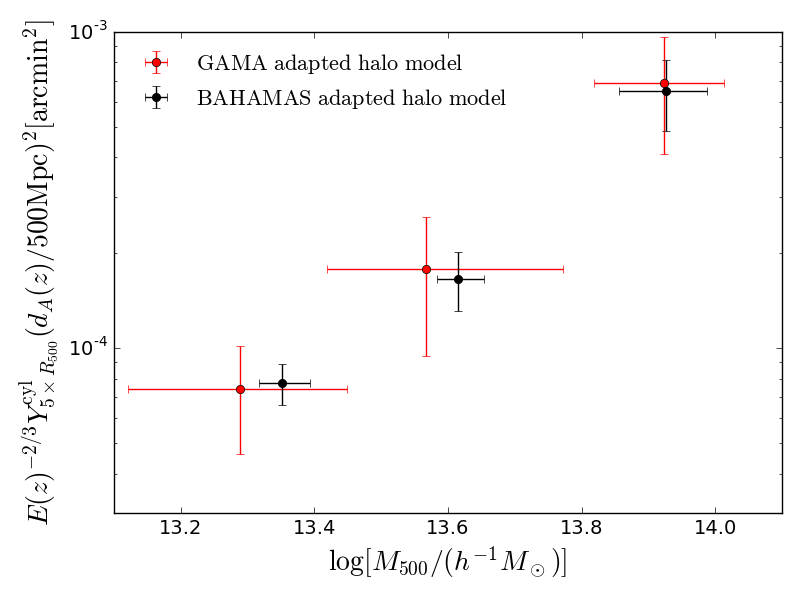}
    }

\caption{X-ray and tSZ scalings with halo mass. {\it Left panel:} Comparison of the observed and predicted soft band X-ray luminosity $L_{X,500}-M_{500}$ relations.  {\it Right panel:} Comparison of the stacked $Y_{5\times R_{500}} - M_{500}$ relations.  In both panels the observed galaxy groups of the GAMA survey are presented by the red data points and the BAHAMAS simulations with the black data points.  The halo masses shown are those derived from the more flexible adapted halo model of \citet{Viola2015a}.}\label{fig:LxYM_adaptedMass}
\end{figure*}

We now combine the measurements of the lensing masses, X-ray luminosities, and tSZ effect fluxes to derive the scalings between the hot gas content and total mass of groups.  We show the $L_X-M_{500}$ and $Y^{cyl}_{5\times R_{500}}-M_{500}$ relations of the GAMA and BAHAMAS galaxy groups in the left and right panels of Figure \ref{fig:LxYM_adaptedMass} respectively.  The masses used here are based on the adapted halo model described in Section \ref{sec:Lensing}.


The results are largely consistent with those of Fig.~\ref{fig:BaryonOnly}, in the sense that there is excellent concordance between the observed and simulated scalings involving the tSZ flux, but also that there is an amplitude mismatch in the scaling involving X-ray luminosity as was shown in the middle panel of Fig.~\ref{fig:BaryonOnly}. Broadly speaking, it appears that the simulation provides an excellent description of the overall gas and stellar content of the groups, but it does not reproduce in detail the central regions (from which the vast majority of the X-ray luminosity originates) of this optically-selected group sample.

\citet{McCarthy2017} found a similar offset between this simulation and the observed X-ray luminosity scalings with stellar mass and stacked weak lensing mass of the optically-selected `locally brightest galaxy' sample of \citet{Anderson2015a, Wang2016} (see fig.~22 of \citealt{McCarthy2017}).  However, no such offset was seen in their comparison with the observed X-ray luminosity$-$halo mass relation of {\it X-ray-selected} groups (see fig.~16 of \citealt{McCarthy2017}).

We note that we have corrected for the effects of Galactic absorption, something that is not present in the simulations. However this only increases the luminosities by at most $\sim 15$ percent, whereas the offset between data and simulations is closer to a factor of $2-3$.  We therefore conclude that this effect is not significant enough to reconcile these offsets. The most plausible explanation is therefore that observed X-ray-selected groups are somewhat biased in terms of their mean X-ray luminosities and that the feedback in the simulations is still not sufficiently energetic in the central regions of groups and clusters.  An interesting future challenge for the feedback modelling, therefore, is to see if it is possible to simultaneously match the overall gas and stellar fractions of optically-selected groups while also reproducing, in a detailed sense, their radial gas distributions in the central regions.

\begin{figure*}
\centering

\subfloat{\includegraphics[width=8.7cm]{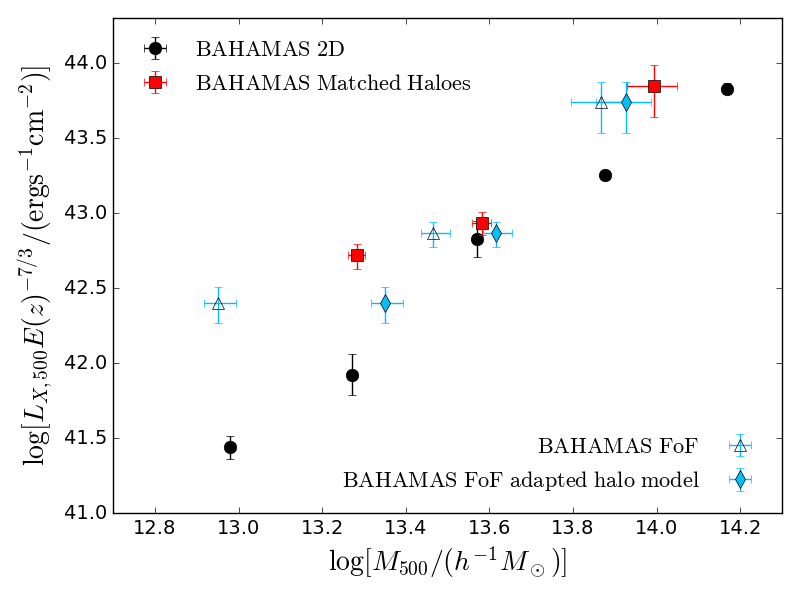}}
~
\subfloat{\includegraphics[width=8.7cm]{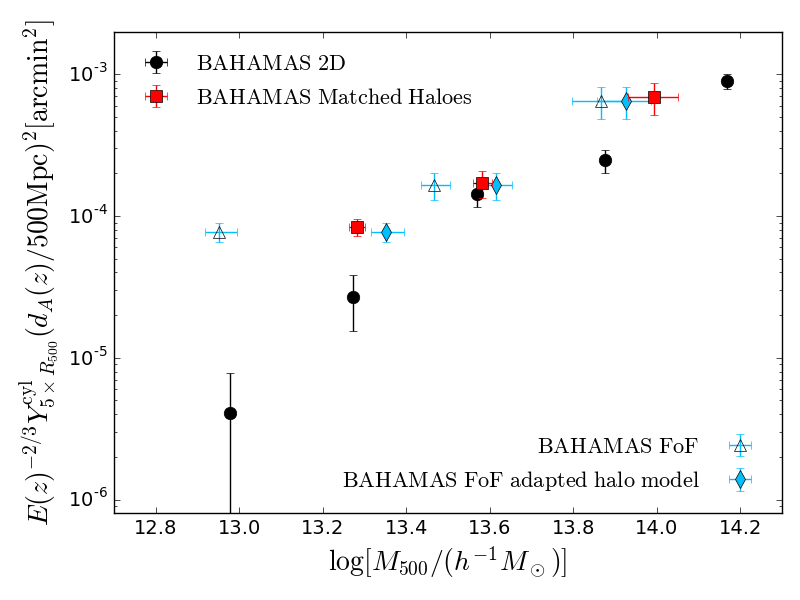}}

\caption{The effects of fragmentation and aggregation on the $L_{X,500}-M_{500}$ ({\it left}) and $Y_{5\times R_{500}} - M_{500}$ ({\it right}) relations.  The black points show the intrinsic truth from 2D simulations (as Fig. \ref{fig:2Dvs3D}); the open triangles show the scaling relations using the fiducial halo model for the lensing mass; the filled diamonds as the triangles but with the adapted model and the red squares are FoF groups re-centred on the central galaxy of the matched group. See Section 6 for details on the matching.}\label{fig:FoFvsTrueHaloes}
\end{figure*}

\section{Selection effects of the Friends-of-Friends algorithm}\label{fragmentation}

In Section 3.2 and Appendix \ref{FoFvsTrue} we discussed the performance of the FoF group finder by comparing the recovered and true group catalogues for BAHAMAS, concluding that fragmentation of massive groups/clusters occurs.  While such fragmentation does not inhibit our ability to compare the simulations and observations, since both were subjected to the same group identification procedure, it does affect our ability to recover the true hot gas$-$halo mass relations. Here we investigate the effects that fragmentation and aggregation of galaxy groups by the FoF algorithm have on the recovered scaling relations. 
In order to do this, we compare the scaling relations found from the FoF analysis of the simulations to the underlying true simulation relation and also create a synthetic relation by matching the FoF groups to the true simulation groups.

Fig. \ref{fig:FoFvsTrueHaloes} shows a summary of our findings. In both panels, the black points are the same as the black points shown in Fig. \ref{fig:2Dvs3D}. These points represent the "truth" established by the simulation. The scaling relations obtained from the FoF analysis, identified by the other points on figure \ref{fig:FoFvsTrueHaloes}, should be compared to this "truth", which, ideally, they should recover.
The solid blue diamonds correspond to the BAHAMAS scaling relations found from our Friends-of-Friends analysis. Here, the adapted HOD modelling described in section \ref{sec:Lensing} was used. 
The open blue triangle points show the relations obtained if we use the fiducial (or standard) HOD modelling instead. The fiducial model is clearly unable to recover the true scaling relation and while the adapted model is much closer to the black points, it does not recover them. The discrepancy between these FoF-based scaling relations and the 'true' relation cannot be caused by the stacking analysis itself since we demonstrated in Section \ref{sec:2Dvs3D}) that this procedure is unbiased. 

We show below that this is likely caused by the fragmentation and aggregation of the groups identified by the FoF finder as these lead to two effects that are likely to impact the scaling relations: The first effect is that the group centre assigned by the FoF will not be the true (optical) centre of the group\footnote{Note that the mis-centring due to fragmentation and aggregation is separate issue from the problem that the central galaxy might not trace the centre of the matter distribution.}, which will cause some of the aperture X-ray and tSZ fluxes to be underestimated. The second effect is that the group stellar mass will be under- or overestimated because of missing members or interlopers, leading to groups ending up in the wrong stellar mass bin.

In order to investigate the impact of the above two effects
we now stack the Xray and tSZ signal using a matched group catalogue. In this catalogue, which is discussed in detail in appendix \ref{FoFvsTrue}, each BAHAMAS FoF group is matched to the most likely true group in the simulation. The matched version of every FoF group now uses the correct centre for the aperture defined by the central galaxy of the matched (true) group. The halo mass of each stack is then defined as the mean halo mass of these matched haloes in each stellar mass bin. The result is shown in Fig.~\ref{fig:FoFvsTrueHaloes} as the red squares. 
We find that the re-centring of the aperture has little effect on the amplitude of the tSZ signal, whereas the X-ray signal increases significantly in the lowest mass bin. This is caused by the fact that the X-ray flux is strongly peaked around the true centre, therefore a wrong centre, due to e.g. fragmentation of the FoF selected groups, could lead to missing a significant part of the signal.
Moreover, the mis-centring can also result in over-subtraction of the background, which would even amplify the previous effect leading to an underestimation of the X-ray luminosity of the blue points (triangles and diamonds). As expected, the re-centring of the aperture has the strongest effect in the first bin, where the fragmentation is strongest. 

The matched groups that constitute the red squares sample in Fig.~\ref{fig:FoFvsTrueHaloes} are a subset of the true halo catalogue from which the true relation (black points) was generated. The only quantity that enters into the red squares that stems from the FoF analysis is the group stellar mass on basis of which the 'matched groups' have been assigned to a stellar mass bin. Given that the red squares do not align with the scaling relation traced by the black points, the logical conclusion is that the groups were assigned to the wrong stellar mass bin, thus causing halo masses to be mixed stronger between the different bins than might be reasonably expected from the scatter in the $M_*^\mathrm{grp}-M_\mathrm{halo}$ relation. The mixing of halo masses between bins is further investigated in Appendix \ref{HaloMassMapping} where we re-map the groups from the stellar mass bins back to their original value and we indeed recover the correct (black points) scaling relations.

This leads to the conclusion that fragmentation/aggregation of the FoF groups finder is responsible for the deviation of the scaling relations. It is caused by two effects combined: the mis-centring of the apertures which causes an underestimation of tSZ- and in particular X-ray-flux, and the mixing of halo masses between different bins. The first former causes the data points to shift downward on the $L_X-M$ plain and to a lesser extent the $Y-M$ plain, compared to where they need ought to be in case of correctly centred apertures. This is captured in the difference (in $L_X$-/$Y$-values) between the red squares and blue diamonds in figure \ref{fig:FoFvsTrueHaloes}. The latter effect however, causes slightly more non-trivial shifts on the aforementioned plain, which we can illustrate with the use of an example. Imagine a massive cluster that ends up in a low (stellar-) mass bin due to fragmentation by the FoF finder. This cluster will increase the mean mass of the stack slightly causing a slight rightward shift \footnote{Compared to the position it would have been had there only been low mass systems in the bin as one naively would expect based on the $M_*^\mathrm{grp}-M_\mathrm{halo}$ relation from the simulations.} in figure \ref{fig:FoFvsTrueHaloes}. However, it will increase the X-ray luminosity or tSZ value even more due to the steep scaling of these quantities with halo mass, causing the stacked data point to shift upwards in figure \ref{fig:FoFvsTrueHaloes} by a greater amount and to move away from the underlying relation (the black points in the same figure). In appendix \ref{HaloMassMapping} we show that if we rebin all groups from their FoF-stellar mass bin into halo mass bins, we indeed recover the underlying scaling relations. 

\section{Discussion and Summary}\label{Discussion}
In this paper we presented a consistent comparison of the stacked weak lensing calibrated X-ray luminosity- and tSZ - mass scaling relations of FoF galaxy groups between the GAMA galaxy survey and the BAHAMAS hydrodynamical simulation. To do so, we defined an approximately volume limited ($z <0.2$) sample of groups  180 deg$^2$ and 625 deg$^2$ of constructed light cones, of the GAMA data and BAHAMAS simulation respectively and compared their properties. 

We showed that we can select similar groups in both observations and simulation, resulting in statistically equivalent systems; this is essential for any meaningful comparison. With the selected group samples we show that the BAHAMAS simulation reproduce the $Y-M_*$ and $Y-M_{500}$ relations strikingly well, which is a direct indication that the integrated electron pressure, and hence the density and temperature distributions, are realistic in the simulation. The overall scaling of the X-ray luminosity with both stellar and halo mass agrees well between data and simulation, but the BAHAMAS results show a overall amplitude offset with respect to the observations.  As \cite{McCarthy2017} have shown that the simulations reproduce the X-ray luminosities of X-ray-selected systems, the offset here with respect to our optically-selected sample may suggest that X-ray selection results in a biased subset of the group population (e.g., \cite{Anderson2015a}) and that the simulations may require somewhat more efficient late-time feedback to reproduce the X-ray luminosities of an unbiased sample.

We measure the excess surface mass density profiles from the gravitational shear induced by the group in both the simulation and around the GAMA groups using the weak lensing data from KiDS and find that the BAHAMAS matter density profiles recover the observational data well. We find that our approach to mimic the observations is robust against projection effects and other systematics.

From the comparison of the recovered scaling relations from our FoF analysis with the underlying scaling relations in the simulation, we find that our group catalogues suffer from serious fragmentation and aggregation issues. This leads to mixing of objects in the relations between stellar-mass and halo-mass relation, which in turn leads to biases in the $Y-M$ and $L_X-M$ relations, and consequently, an offset in the amplitude and a flattening of the slope in both the tSZ and X-ray luminosity relations. We investigated the fragmentation and aggregation by matching the FoF groups to the underlying simulation groups and conclude that this selection effect does not affect our comparison between data and simulation as they both suffer from it in the same way, although it does prevent us from comparing our FoF analysis directly to \cite{Wang2016}.

The fact that the recovered scaling relations of the GAMA and BAHAMAS FoF groups agree so well, despite the problems with the FoF algorithm, is the consequence of the realism of the simulation and our ability to select equivalent samples in both observations and data. It also highlights the need for realistic hydrodynamical simulation to test our assumptions and correctly interpret observational data. 
Being able to make such comparisons for galaxy group samples is an important step to validate the use of hydrodynamical simulation to better understand the baryon physics governing the structure of these systems and hence the matter distribution at small scales. 

This does not take away the point that in order to fully understand these structures in future surveys, and the relation to baryon physics, it will be necessary to mitigate these selection biases altogether, particularly when studying the hydrostatic structure of these groups via the stacked density profiles. We note that issues with fragmentation/aggregation potentially affect any group finding technique, implying that this is an issue that must be investigated systematically, which can now be done with the aid of hydrodynamical simulation.

To fully exploit the power of groups to constrain baryon physics, larger samples will also be required. We are currently working towards resolving problems with simulating the optical luminosity function in order to unlock the full flux limited GAMA group sample. This would tackle two problems at once: firstly the sample size would increase significantly and secondly, the full luminosity dependent GAMA Friends-of-Friends group finder \citep{Robotham2011} can be used to more robustly select galaxy group in both data and simulation.

\section*{Acknowledgements}\label{sec:Acknowledge}
We would like to thank Thomas Reiprich for useful input on x-ray counts to flux conversion and Alex Mead for useful discussion. LvW, GH and TT are supported by NSERC and CIfAR Cosmology and Gravity program. HH acknowledges support from Vici grant 639.043.512, financed by the Netherlands Organisation for Scientific Research (NWO). MV acknowledges support from the European Research Council under FP7 grant number 279396 and the Netherlands Organisation for Scientific Research (NWO) through grants 614.001.103.  CH acknowledges support from the European Research Council under grant number 647112.  H. Hildebrandt is supported by an Emmy Noether grant (No. Hi 1495/2-1) of the Deutsche Forschungsgemeinschaft.KK acknowledges support by the Alexander von Humboldt Foundation. JS acknowledges the support from the Netherlands Organization for Scientific Research (NWO) through the VICI grant 639.043.409. EvU acknowledges support from an STFC Ernest Rutherford Research Grant, grant reference ST/L00285X/1. GVK acknowledges financial support from the Netherlands Research School for Astronomy (NOVA) and Target. Target is supported by Samenwerkingsverband Noord Nederland, European fund for regional development, Dutch Ministry of economic affairs, Pieken in de Delta, Provinces of Groningen and Drenthe.

Based on observations made with ESO Telescopes at the La Silla Paranal Observatory under programme ID 179.A-2004. 

GAMA is a joint European-Australasian project based around a spectroscopic campaign using the Anglo-Australian Telescope. The GAMA input catalogue is based on data taken from the Sloan Digital Sky Survey and the UKIRT Infrared Deep Sky Survey. Complementary imaging of the GAMA regions is being obtained by a number of independent survey programmes including GALEX MIS, VST KiDS, VISTA VIKING, WISE, Herschel-ATLAS, GMRT and ASKAP providing UV to radio coverage. GAMA is funded by the STFC (UK), the ARC (Australia), the AAO, and the participating institutions. The GAMA website is \url{http://www.gama-survey.org/} . 

This research has made use of data and webtools obtained from NASA’s High Energy Astrophysics Science Archive Research Center (HEASARC), a service of Goddard Space Flight Center and the Smithsonian Astrophysical Observatory. The data we used has been provided in the convenient \texttt{HEALPix} format \citep{Gorski2005} by the Centre d'Analyse de Données Etendues (CADE) or Analysis Center for extended data.

{\it Author Contributions}: All authors contributed to the development and
writing of this paper. The authorship list reflects the lead authors of this paper (AJ, MV, IMcC, LvW, HH and AR), followed by two alphabetical groups. The first alphabetical group (GH, AH, HT and TT) consists of authors who contributed to the analysis itself. Members of the second group are infrastructure contributors whose products are directly used in this work, and/or have contributed to the writing.


\bibliography{library}

\appendix
\section{FoF versus true groups}\label{FoFvsTrue}
To investigate the possible selection effects that arise from our choice of a fixed linking length, we match all the BAHAMAS FoF groups to the true simulation groups based on galaxy membership. For this matching we follow the procedure outlined in \citet{Robotham2011} and define the most probable match to be the one that maximises
\begin{align}
	P_\mathrm{FoF} &= \frac{N_\mathrm{overlap}^2}{N_\mathrm{FoF}N_\mathrm{true}}\ ,
\label{Pfof}
\end{align}
where $N_\mathrm{FoF}$ is the multiplicity of the FoF group, $N_\mathrm{true}$ is the multiplicity of the true group and $N_\mathrm{overlap}$ is the number of galaxies in common between the FoF and the true group. 

As an example, imagine there is a FoF group with 6 members, 4 of which are in common with a true group of 5 members and 2 of which are in common with a true group of 9 members. According to the above metric the most probable match is the true group having 4 members in common, for which $P = (4\times4)/(6\times 5) = 0.53$ compared to $P = (2\times2)/(6\times9) = 0.07$.

We find that only 63 percent of the FoF groups are bijectively matched to a true halo, where the criterion for a bijective match is that the joint population of the true and the FoF group includes more than 50 percent of their respective members. The remaining 37 percent of the groups are either matched to more than 1 true group (aggregation), or multiple FoF groups are matched to the same true group (fragmentation).

\begin{figure}
\centering
\includegraphics[width=8.7cm]{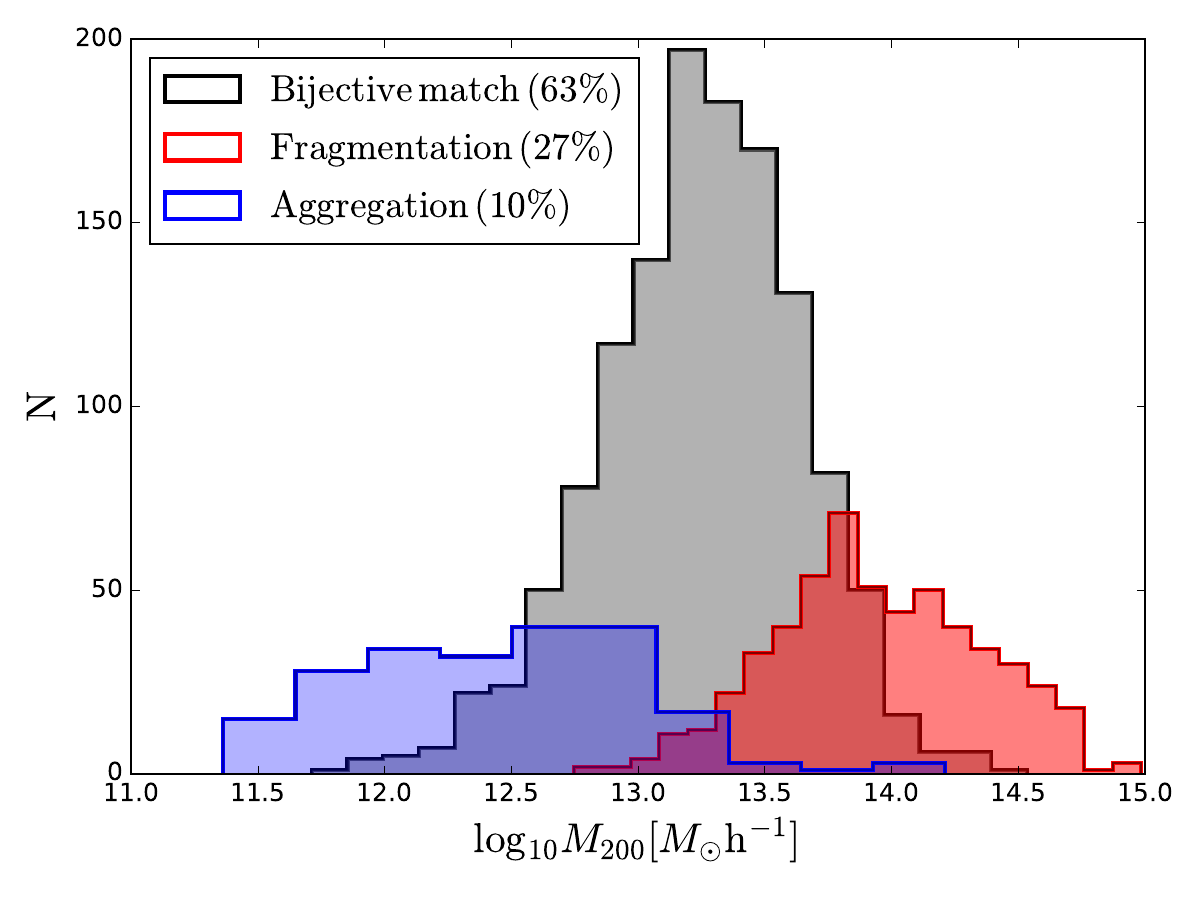}
\caption{Distribution of total halo mass for true groups that are bijectively matched to FoF groups (black), ones that have been fragmented by the FoF algorithm (red) and the ones that have been aggregated by the FoF algorithm (blue).}\label{fig:fragmention_fractions}
\end{figure}

Having investigated the matching of FoF groups to true groups more closely, we find that the aggregation/fragmentation is roughly a function of halo mass; we find that most of the highest mass haloes, i.e.  $\log[M_{200}/(h^{-1}M_\odot)] > 14 $,  are fragmented into sub-haloes whereas the low mass systems are mostly aggregated groups, see Figure \ref{fig:fragmention_fractions}. This has the effect that the assigned group stellar mass, i.e. the sum of the stellar masses of the FoF group members is grossly underestimated for high mass haloes and these can populate lower stellar mass bins, possibly multiple times; for every FoF group linked to this halo. On the other side, aggregated FoF groups can dilute the higher stellar mass bins, although the effect of this is less severe due to the steep scaling of the X-ray luminosity and tSZ signal with halo mass. 

\begin{figure}
\centering
\includegraphics[width=8.7cm]{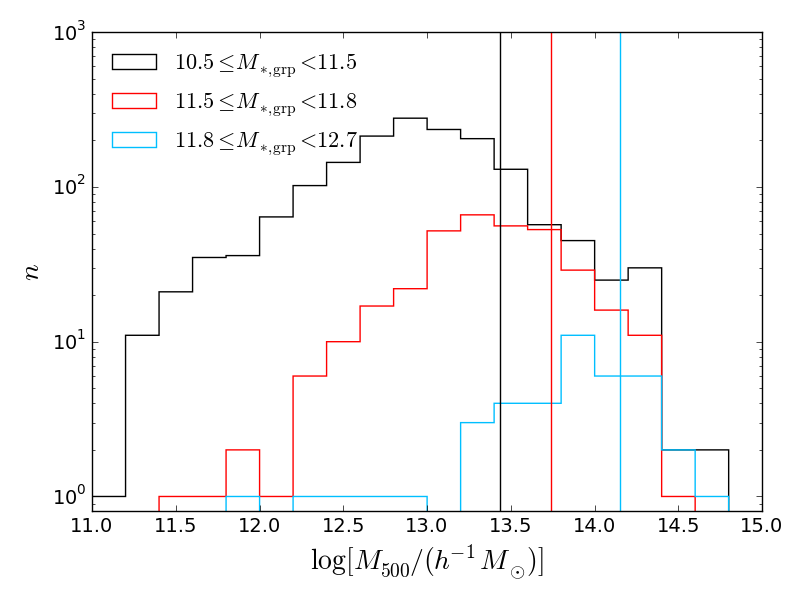}
\caption{The halo masses of the haloes matched to the FoF groups in the three stellar mass bins. The solid vertical lines show the mean in each bin. We see that due to fragmentation of large haloes by the FoF algorithm, the lowest stellar mass bin contains haloes spanning the entire range up.}\label{fig:HaloMassesInBins}
\end{figure}

Figure \ref{fig:HaloMassesInBins} shows, for the matched groups, how the true halo masses are indeed distributed in a given group stellar mass bin: we can clearly see that the fragmentation and aggregation effect contaminates group stellar mass bins.
All three bins contain haloes of a wide range in masses, but the fragmentation of massive haloes clearly has the strongest effect in the lowest stellar mass bin, which contains, against a naive expectation based on the $M_*^\mathrm{grp}-M_\mathrm{halo}$ relation from simulations, many massive haloes. In section \ref{fragmentation} we discuss the effects this has on the scaling relations. 

\section{Halo model fits}\label{HaloModelFits}
Here we list all the best fit parameters used in this study. Tables \ref{table:simplemodel} and \ref{table:adaptedmodel} shows the  fit with the 68 percent confidence interval of the standard halo model parameters as described in \citep{Viola2015a} (however, we warn against reading too much into their physical meaning given the aggregation/fragmentation issues). The meaning of the parameters are described in Section 4.1: $f_c$, $\sigma_c$, $log(M^{1,2,3})$, $p_{\rm off}$ and $r_{\rm off}$ correspond to the amplitude of the NFW mass-concentration relation, the width of the log-normal HOD, the mean of the log-normal HOD for the three bins, the probability for the BCG to be mis-centred, and the amount of mis-centring from the bottom of the gravitational potential well.



\begin{table*}
\caption{Halo model parameters fitted to the simulations (top table) and data (botton table) for the simple model case. A full description of the parameters are given in \citep{Viola2015a}. \label{table:simplemodel}}
\begin{tabular}{ccccccc}
& & & BAHAMAS & & & \\
\hline
$f_\mathrm{c}$ & $\sigma_\mathrm{c}$ & $p_\mathrm{off}$ & $r_\mathrm{off}$ & $\log[\frac{M_{200}^1}{h^{-1}\mathrm{M_\odot}}]$ & $\log[\frac{M_{200}^2}{h^{-1}\mathrm{M_\odot}}]$ & $\log[\frac{M_{200}^3}{h^{-1}\mathrm{M_\odot}}]$ \\
\hline
$0.64_{-0.02}^{+0.02}$ & $0.70_{-0.04}^{+0.03}$ & $0.27_{-0.25}^{+0.43}$ & $0.02_{-0.01}^{+0.05}$ & $13.14_{-0.02}^{+0.03}$ & $13.67_{-0.03}^{+0.03}$ & $14.09_{-0.05}^{+0.05}$ \\
\hline
\\
& & & GAMA & & & \\
\hline
$f_\mathrm{c}$ & $\sigma_\mathrm{c}$ & $p_\mathrm{off}$ & $r_\mathrm{off}$ & $\log[\frac{M_{200}^1}{h^{-1}\mathrm{M_\odot}}]$ & $\log[\frac{M_{200}^2}{h^{-1}\mathrm{M_\odot}}]$ & $\log[\frac{M_{200}^3}{h^{-1}\mathrm{M_\odot}}]$ \\
\hline
$0.66_{-0.13}^{+0.17}$ & $0.56_{-0.18}^{+0.10}$ & $0.34_{-0.24}^{+0.37}$ & $0.20_{-0.14}^{+0.21}$ & $13.18_{-0.10}^{+0.10}$ & $13.66_{-0.08}^{+0.08}$ & $14.17_{-0.10}^{+0.09}$ \\
\hline
\end{tabular}
\end{table*}

\begin{table*}
\caption{Halo model parameters fitted to the simulations (top table) and data (bottom table) for the adapted model case. Superscripts $^{1, 2, 3}$ correspond to the three mass bins defined in Table \ref{table:binningschemes_GAMA}. A full description of the parameters are given in \citep{Viola2015a}. \label{table:adaptedmodel}}
\begin{tabular}{ccccccccccccccc}
& & & & BAHAMAS & & & \\
\hline
$f_\mathrm{c}^1$ & $f_\mathrm{c}^2$ & $f_\mathrm{c}^3$ & $\sigma_\mathrm{c}^1$ & $\sigma_\mathrm{c}^2$ & $\sigma_\mathrm{c}^3$ & $p_\mathrm{off}^1$ & $p_\mathrm{off}^2$ & $p_\mathrm{off}^3$ \\
\hline
$2.28_{-0.23}^{+0.16}$ & $2.24_{-0.26}^{+0.18}$ & $0.49_{-0.09}^{+0.09}$ & $0.81_{-0.14}^{+0.12}$ & $0.78_{-0.13}^{+0.10}$ & $0.49_{-0.09}^{+0.11}$ & $0.75_{-0.02}^{+0.02}$ & $0.71_{-0.04}^{+0.03}$ & $0.00_{-0.00}^{+0.00}$ \\
\hline
\hline
$r_\mathrm{off}^1$ & $r_\mathrm{off}^2$ & $r_\mathrm{off}^3$ & $\log[\frac{M_{200}^1}{h^{-1}\mathrm{M_\odot}}]$ & $\log[\frac{M_{200}^2}{h^{-1}\mathrm{M_\odot}}]$ & $\log[\frac{M_{200}^3}{h^{-1}\mathrm{M_\odot}}]$ \\
\hline
$5.38_{-0.47}^{+0.40}$ & $3.13_{-0.38}^{+0.34}$ & $4.54_{-2.93}^{+3.36}$ & $13.62_{-0.04}^{+0.04}$ & $13.87_{-0.03}^{+0.03}$ & $14.34_{-0.06}^{+0.05}$ \\
\hline
\\
& & & & GAMA & & & \\
\hline
$f_\mathrm{c}^1$ & $f_\mathrm{c}^2$ & $f_\mathrm{c}^3$ & $\sigma_\mathrm{c}^1$ & $\sigma_\mathrm{c}^2$ & $\sigma_\mathrm{c}^3$ & $p_\mathrm{off}^1$ & $p_\mathrm{off}^2$ & $p_\mathrm{off}^3$ \\
\hline
$1.38_{-0.72}^{+0.75}$ & $1.63_{-0.57}^{+0.59}$ & $0.57_{-0.20}^{+0.29}$ & $0.71_{-0.28}^{+0.20}$ & $0.65_{-0.27}^{+0.20}$ & $0.40_{-0.23}^{+0.18}$ & $0.64_{-0.25}^{+0.15}$ & $0.52_{-0.24}^{+0.19}$ & $0.00_{-0.00}^{+0.00}$ \\
\hline
\hline
$r_\mathrm{off}^1$ & $r_\mathrm{off}^2$ & $r_\mathrm{off}^3$ & $\log[\frac{M_{200}^1}{h^{-1}\mathrm{M_\odot}}]$ & $\log[\frac{M_{200}^2}{h^{-1}\mathrm{M_\odot}}]$ & $\log[\frac{M_{200}^3}{h^{-1}\mathrm{M_\odot}}]$ \\
\hline
$4.81_{-2.45}^{+3.00}$ & $5.42_{-3.82}^{+3.02}$ & $4.40_{-2.88}^{+3.43}$ & $13.57_{-0.18}^{+0.23}$ & $13.88_{-0.16}^{+0.24}$ & $14.32_{-0.13}^{+0.13}$ \\
\hline
\end{tabular}
\end{table*}

\section{Bright X-ray sources}\label{Xpointsources}
When studying X-ray emission from the thermal Bremsstrahlung from the hot intracluster medium (ICL) one needs to consider the possible contribution from other X-ray sources, such as compact objects like X-ray binaries and active galactic nuclei (AGN). While the contribution of the former is shown in \citet[Fig. 5]{Anderson2015a} to be negligible for the mass range of the systems in our studies, the latter source of X-ray emission requires more careful approach. 

The X-ray luminosity in the simulations comes only from the hot gas content of the haloes. Known AGN X-ray sources should therefore be masked in a direct comparison between observations and simulation data. We investigated the effect of masking bright sources $> 2\times 10^{-2} \mathrm{photon\ counts/s}$ listed in the Second ROSAT All-Sky Survey (2RXS) source catalogue \citep{Boller2016} and found, in line with the findings of \citet{Anderson2015a}, that masking these sources has a relatively small effect on the total stacked flux. Reducing the flux within the different bins by a small fraction $\leq 20\% $ depending on the bin.

We note that we are dealing with a very low redshift sample of groups so that most of the AGN listed in the in the 2RXS source catalogue will lie at a higher redshift and therefore are not correlated with the GAMA groups in our sample and mainly add to the background, which is subtracted in our stacking analysis. We therefore decided not to remove point sources listed in the 2RXS source catalogues from the RASS data when stacking. The possible flux bias we introduce this way is $< 10\%$ in the two lower mass bins and $\sim 20\% $ in the highest mass bin, which is within the errorbars of the X-ray luminosities of our sample and hence will not affect the conclusions of our study.

\section{Counts to Flux conversion}\label{sec:counts2flux}
As the flux levels of most faint X-ray sources are very low, observational X-ray data is commonly provided in photon counts per unit time within an energy band. These {\it count-rates} are then converted into an energy flux based on assumptions of the source's rest frame spectrum, the galactic gas column density and the telescope's efficiency over the observed energy range (this is provided in the response matrix of the telescope). 
	
In this work we convert the photon count-rate of the GAMA groups into an energy flux using the webPIMMS\footnote{\url{http://heasarc.gsfc.nasa.gov/cgi-bin/Tools/w3pimms/w3pimms.pl}} tool provided by NASA's High Energy Astrophysics Science Archive Research Center (HEASRAC). We select the conversion Rosat PSPC counts to flux, choosing the APEC model \citep{Smith2001} to model the average source spectrum for a stack. The plasma temperature for the model is calculated using the weak lensing calibrated $M_{500} - T$ relation from \citet{Kettula2013}. We assume a plasma metallicity of $Z = 0.4 Z_\odot$ following \citet{Anderson2015a}\footnote{WebPIMMS makes use of pre-calculated APEC models for given metallicity and temperature (spanning the range 0.0343 keV - 27.25 keV). We choose the model with a plasma temperature closest to the average plasma temperature we calculated for the stacks}. 
	
The model source is placed at the median redshift of the groups within the bin. By providing a source redshift for the model, WebPIMMS includes the $k$-correction corresponding to this redshift in the predicted flux rates. 
	
The Galactic hydrogen column is taken to be $3.0\times 10^{20}\mathrm{cm^{-2}}$, which is the median value for the different sight lines to the groups as calculated from the $N_\mathrm{H}\mathrm{tot}$ tool\footnote{\url{http://www.swift.ac.uk/analysis/nhtot/}} \citep{Willingale2013}. This tool combines the $H_\mathrm{I}$ from the {\it Leiden/Argentine/Bonn} Survey \citep{Kalberla2005} and the dust map from \mbox{\citet{Schlegel1998}}. 
		
WebPIMMS calculates a predicted flux at given count-rate and for a given Galactic gas column density $N_H$, but also provides an unabsorbed flux estimate which corrects for the scatter and absorption due to the galactic gas column density provided. Since we are interested in this unabsorbed flux we use the latter conversion factor for our conversion from flux to luminosity. 
The counts to flux factors $C_\mathrm{flux}$ are provided in Table \ref{table:results_GAMA}; note that these conversion factors already include the appropriate $k$-correction. 
    	
As the X-ray maps from the BAHAMAS simulations are given in observer frame flux, we only $k$-correct these fluxes back into the rest-frame using the ratio of a APEC model at the median group redshift of a stack to that same model at redshift 0. The $k$ correction factors are listed in Table \ref{table:results_BAHAMAS}.

\section{Null tests for X-ray and tSZ stacking}\label{NullTest}
\begin{figure}
\centering
\includegraphics[width=8.7cm]{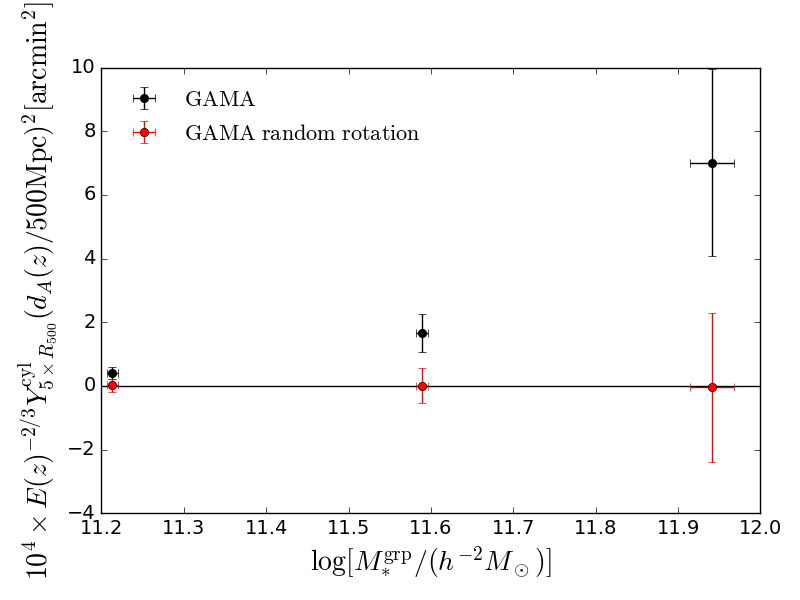}
\includegraphics[width=8.7cm]{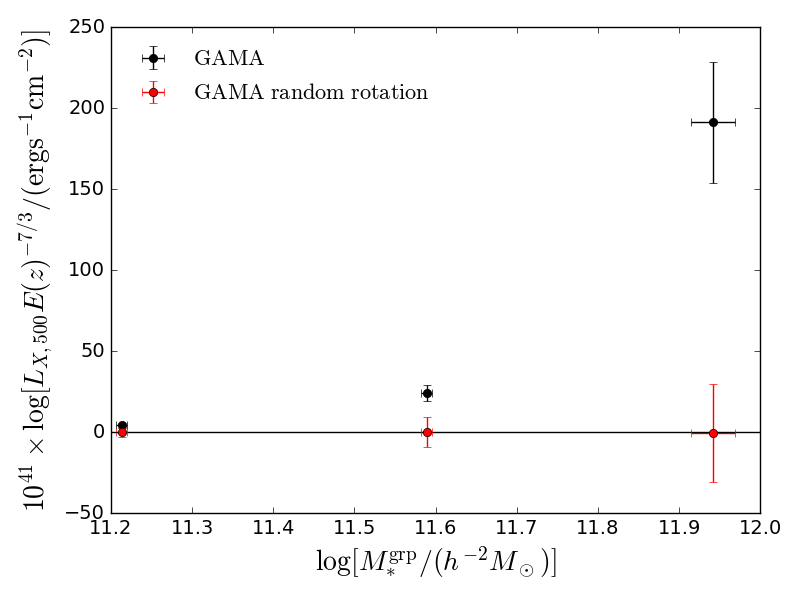}
\caption{{\it Top panel:} The stacked $ Y_{5\times R_{500}} - M_\mathrm{*,grp}$ relation for the GAMA galaxy groups (black) and mean of 1000 realisations of the GAMA groups randomly rotated along their galactic latitude (red). {\it Bottom panel:} Similar comparison, but for the X-ray luminosity. In both cases we see that the signal vanishes when we randomize the positions of the groups.
\label{fig:Random}}
\end{figure}
In this section we show that our X-ray and tSZ results are robust against systematic errors introduced by our stacking procedure. For a similar an analysis of the lensing measurement we refer the reader to \citet{Viola2015a}. 

We perform the same stacking procedures outlined in Section~\ref{sec:DataStacking} but this time we randomise the Galactic longitude of the groups\footnote{We rotate along the longitudinal direction instead of a complete randomisation of the group's position to prevent it from moving into the galactic plane.} so that it effectively becomes a random measurement. We keep the group's redshift and stellar mass the same. We repeat this random stacking 1000 times and measure the mean and 1$\sigma$ errors of these 1000 realisations, and present the results in Fig.~\ref{fig:Random}.
The top and bottom panels show the comparison of the random and real $Y-M_*$ and $L_X-M_*$ relations respectively. We conclude that our measurement is has no significant systematic bias.

\section{Halo mass mixing}
\begin{figure}
\centering
\includegraphics[width=8.7cm]{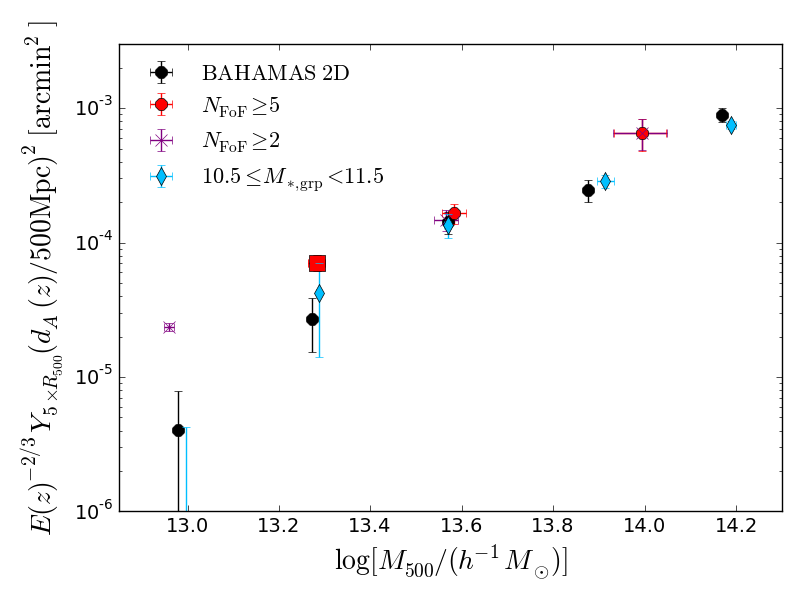}
\includegraphics[width=8.7cm]{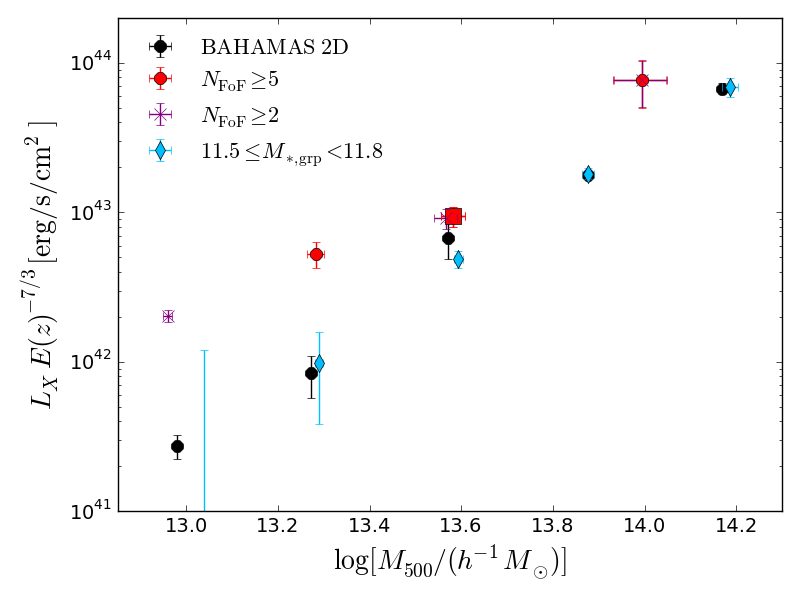}
\caption{The $Y-M$ ({\it top}) and $L_X-M$ ({\it bottom}) relation of the BAHAMAS galaxy groups. The black points show the relations of the true groups measured in 2D and binned according to their halo mass. The red points and purple crosses show FoF relations binned in stellar mass with and without the apparent richness cut  $N_\mathrm{FoF}\geq 5$. The blue points are the groups that constitute the first (top panel) and second (bottom panel) stellar mass bins (red points), but have been rebinned according to their true halo mass. The large red squares highlight the stellar mass bins in which the blue points fall based on their FoF stellar mass.} \label{fig:FoF_Fragmentation}
\end{figure}
In section \ref{fragmentation} we have shown that the stacks of Friends-of-Friends groups suffer from a mixing of halo masses. This is due to large haloes being fragmented into smaller groups and small haloes being aggregated into larger groups. In this appendix we show how the true BAHAMAS haloes matched to each FoF group in a given stellar mass bin map on to the real $Y-M$ and $L_X-M$ relation. We also investigate the effect of the apparent richness cut imposed in our selection of FoF groups. 

In figure \ref{fig:FoF_Fragmentation} we compare the BAHAMAS $Y-M$ (top) and $L_X-M$ (bottom) scaling relations with (red) and without (purple) the multiplicity cut $N_\mathrm{FoF}\geq 5$. Plotted are the stacks of the groups matched to the FoF groups as described in section \ref{Results}. We find, as expected from figure \ref{fig:HaloMassesInBins}, the lowest stellar mass bin drops in both average halo mass and its gas hot gas content. The multiplicity cut, however, does not seem affect the slope of the overall scaling relations. And the shallow slope can be completely contributed to the halo mass mixing due to fragmentation and aggregation of galaxy groups. 

To more directly illustrate the effect the mixing of halo masses has on the scaling relations we show, in the top panel of figure \ref{fig:FoF_Fragmentation}, the $Y-M$ scaling of the groups in the first stellar mass bin (highlighted as the large red square), when we rebin them according to their true halo mass from the simulations (blue). The black points show the underlying 'true' relation for reference. We see that by 'undoing' the halo mass mixing we recover the underlying scaling relation. The bottom panel of figure \ref{fig:FoF_Fragmentation} shows the same but for the second stellar mass bin and the X-ray luminosity. 
\label{HaloMassMapping}

\end{document}